\documentclass{emulateapj}
\renewcommand{\farcs}     {\mbox{\ensuremath{.\!\!^{\prime\prime}}}}

\renewcommand{\ion}[2]{#1$\;${\scshape{#2}}}

\renewcommand{\deg}{\mbox{$^{\circ}$}}

\newcommand{\fluxunits}{\mbox{erg cm$^{-2}$ s$^{-1}$ \AA$^{-1}$}}
\newcommand{\fluxunitsalt}{\mbox{erg cm$^{-2}$ s$^{-1}$}}

\newcommand{\Msun}{\mbox{$M_{\sun}$}}

\newcommand{\heiiw}{\mbox{\ion{He}{ii} $\lambda$4686}}

\newcommand{\oiaw}{\mbox{[\ion{O}{i}] $\lambda$6300}}

\newcommand{\oiiw}{\mbox{[\ion{O}{ii}] $\lambda$3727}}

\newcommand{\niiw}{\mbox{[\ion{N}{ii}] $\lambda \lambda$6548, 6583}}

\newcommand{\siiw}{\mbox{[\ion{S}{ii}] $\lambda \lambda$6716, 6731}}

\newcommand{\oiiiaw}{\mbox{[\ion{O}{iii}] $\lambda$4959}}
\newcommand{\oiiibw}{\mbox{[\ion{O}{iii}] $\lambda$5007}}
\newcommand{\oiiicw}{\mbox{[\ion{O}{iii}] $\lambda$4363}}
\newcommand{\oiiiw}{\mbox{[\ion{O}{iii}] $\lambda \lambda$4959, 5007}}

\newcommand{\heii}{\mbox{\ion{He}{ii}}}
\newcommand{\hii}{\mbox{\ion{H}{ii}}}
\newcommand{\oi}{\mbox{[\ion{O}{i}]}}
\newcommand{\oii}{\mbox{[\ion{O}{ii}]}}
\newcommand{\nii}{\mbox{[\ion{N}{ii}]}}
\newcommand{\ha}{\mbox{H$\alpha$}}
\newcommand{\sii}{\mbox{[\ion{S}{ii}]}}
\newcommand{\hb}{\mbox{H$\beta$}}
\newcommand{\hg}{\mbox{H$\gamma$}}
\newcommand{\oiii}{\mbox{[\ion{O}{iii}]}}
\newcommand{\civ}{\mbox{\ion{C}{iv}}}

\shortauthors{L.-B. Desroches et al.}
\shorttitle{Multiwavelength Monitoring of NGC\,4395. III.}

\begin{document}


\slugcomment{Accepted for publication in the Astrophysical Journal: June 18, 2006}

\title{Multiwavelength Monitoring of the Dwarf Seyfert 1 Galaxy
NGC\,4395. III. Optical Variability and X-ray/UV/Optical Correlations.}
\author{Louis-Benoit Desroches\altaffilmark{1}, Alexei
V. Filippenko\altaffilmark{1}, Shai Kaspi\altaffilmark{2,3}, Ari
Laor\altaffilmark{2}, Dan Maoz\altaffilmark{3}, Mohan
Ganeshalingam\altaffilmark{1}, Weidong Li\altaffilmark{1}, Edward
C. Moran\altaffilmark{4}, Brandon Swift\altaffilmark{1}, Misty C. 
Bentz\altaffilmark{5}, Luis C. Ho\altaffilmark{6}, Kirpal
Nandra\altaffilmark{7}, Paul M. O'Neill\altaffilmark{7}, and Bradley
M. Peterson\altaffilmark{5}}

\altaffiltext{1}{Department of Astronomy, University of California, Berkeley,
CA 94720-3411; louis@astro.berkeley.edu, alex@astro.berkeley.edu,
mganesh@astro.berkeley.edu, wli@astro.berkeley.edu,
bswift@ugastro.berkeley.edu}
\altaffiltext{2}{Department of Physics, Technion, Haifa 32000, Israel;
laor@physics.technion.ac.il}
\altaffiltext{3}{Wise Observatory and School of Physics and Astronomy, Tel-Aviv
University, Tel-Aviv 69978, Israel; shai@wise.tau.ac.il, dani@wise.tau.ac.il}
\altaffiltext{4}{Department of Astronomy, Wesleyan University, Van Vleck
Observatory, Middletown, CT 06459; ecm@astro.wesleyan.edu}
\altaffiltext{5}{Department of Astronomy, Ohio State University, 140 West 18th
Ave., Columbus, OH 43210-1173; bentz@astronomy.ohio-state.edu,
peterson@astronomy.ohio-state.edu}
\altaffiltext{6}{Carnegie Observatories, 813 Santa Barbara St., Pasadena, CA
91101; lho@ociw.edu}
\altaffiltext{7}{Astrophysics Group, Imperial College London, Blackett
Laboratory, Prince Consort Road, London, SW7 2AZ, UK; k.nandra@imperial.ac.uk,
p.oneill@imperial.ac.uk}


\begin{abstract}
We present optical observations of the low-luminosity Seyfert 1 nucleus of
NGC\,4395, as part of a multiwavelength reverberation-mapping
program. Observations were carried out over two nights in 2004 April at Lick,
Wise, and Kitt Peak Observatories. We obtained $V$-band and $B$-band
photometry, and spectra over the range 3500--6800 \AA. Simultaneous {\it Hubble
Space Telescope} UV and {\it Chandra} X-ray observations are presented in
companion papers. Even though NGC\,4395 was in an extremely low state of
activity, we detect significant continuum variability of 2--10\%, increasing
toward shorter wavelengths. The continuum light curves, both spectroscopic and
photometric, are qualitatively similar to the simultaneous UV and X-ray light
curves. Inter-band cross-correlations suggest that the optical continuum
emission lags behind the UV continuum emission by $24^{+7}_{-9}$ min, and that
the optical continuum emission lags behind the X-ray continuum emission by
$44\pm13$ min, consistent with a reprocessing model for active galactic nucleus
emission. There are also hints of Balmer emission lines lagging behind
the optical continuum by an amount slightly larger than the emission-line lag 
detected in the UV. These results are all similar to those of other Seyfert 1 
nuclei. The emission-line lag yields a mass measurement of the central black 
hole, which although not very significant, is consistent with the value derived 
from the simultaneous UV data.
\end{abstract}



\section{INTRODUCTION}

The nucleus of NGC\,4395 is a ``dwarf'' type 1 Seyfert active galactic nucleus
(AGN), the least luminous known in its class \citep{FS89, FHS93}. Its presence in an
essentially bulgeless, extremely late-type galaxy is surprising; such AGNs (and
quiescent black holes) are usually found in bulge-dominated early-type systems
\citep{HFS97}, with black hole mass and bulge mass strongly correlated
\citep{Magorrian98, Laor98, MH03, HR04}. Despite this, the nucleus of NGC\,4395
exhibits all of the hallmarks of a type 1 AGN: it is detected as a variable
X-ray point source \citep{Lira99, Moran99, Moran05, Iwasawa00, Ho01, SIF03}, it
has a high brightness temperature, highly compact radio core \citep{Moran99,
HU01, WFH01}, and its spectral energy distribution from X-ray to radio
wavelengths is similar to that of other type 1 AGNs \citep{Moran99}.

NGC\,4395 offers a chance to probe the low end of the supermassive black hole
(SMBH) mass distribution. Knowing the mass of the central object will help
constrain the $M_{\mathrm{BH}}-\sigma_*$ relation
\citep{Gebhardt00,Ferrarese01,Tremaine02}, which suggests a strong link between
the growth of the SMBH and evolution of the host galaxy.  The SMBH mass can be measured
through emission-line reverberation mapping \citep{BM82,Peterson93,NP97}.  The
relations of \citet{Kaspi00,Kaspi05} between continuum luminosity and radius of the
broad-line region (BLR) suggest that the size of the Balmer line-emitting
region in NGC\,4395 is of order light hours.

We have used the reverberation method successfully in the ultraviolet (UV) with the {\it
Hubble Space Telescope} ({\it HST}) to determine the size of the BLR and the
mass of the central object, measured to be $M_{\mathrm{BH}} = (3.6 \pm 1.1)
\times 10^5 \ \Msun$ \citep[hereafter Paper I]{Peterson05}. We present here
simultaneous optical observations that complement the UV results. Simultaneous
X-ray observations with {\it Chandra} are presented by \citet[hereafter Paper
II]{ONeill06}. The observations and data reduction are outlined in
Section~\ref{sec:obs}, while the data analysis is described in
Section~\ref{sec:analysis}. In Section~\ref{sec:disc}, we compare our results with those published
for other AGNs and estimate the SMBH mass. We summarize
our results in Section~\ref{sec:con}.

\section{OBSERVATIONS AND REDUCTIONS}
\label{sec:obs}

\subsection{Lick Observatory}

Spectra were obtained on 2004 April 10 and 11 (UT dates are used throughout
this paper) at the Lick Observatory 3-m Shane telescope, using the Kast
spectrograph \citep{MS93}. The instrument separates blue and red light using a
dichroic beamsplitter onto two $1200\times 400$ pixel CCDs (optimized for blue
or red light) for simultaneous coverage. The pixel scale in the
cross-dispersive direction is about $0\farcs 8$ pixel$^{-1}$. The blue-side
grism gave a resolution of 1.83 \AA \ pixel$^{-1}$ with an effective range of
3200--5400 \AA , while the red-side grating was set to give a resolution of
1.17 \AA \ pixel$^{-1}$ with an effective range of 5300--6800 \AA . The slit
width was 2\arcsec. Sequences of 5 to 8 exposures of NGC\,4395, each of
duration 10 min, were bracketed by exposures of standard stars (HZ44 and Feige
34; \citealt{Oke90}). The position angle (PA) of the slit was set to the average parallactic
angle during each sequence to minimize differential light loss
\citep{Filippenko82}.

With BLR variability expected on sub-hour timescales, 10-min exposures were
necessary to maximize temporal coverage while maintaining a reasonable
signal-to-noise ratio (S/N). The standard stars were used to monitor
atmospheric extinction and obtain relative flux calibration; absolute flux
calibration of individual NGC\,4395 spectra was done via narrow emission lines
(expected to be constant in an AGN). The weather was generally clear, though
not photometric, with occasional thin clouds. Conditions worsened slightly on
the second night, with more cirrus clouds. Typical seeing conditions were
variable and ranged from $1\arcsec$ to $2\farcs 5$ full width at half-maximum
intensity (FWHM).

Spectra were bias subtracted, flat fielded, and wavelength calibrated with
standard IRAF\footnote{IRAF is distributed by the National Optical Astronomy
Observatories, which are operated by the Association of Universities for
Research in Astronomy, Inc., under cooperative agreement with the National
Science Foundation.} routines. Wavelength calibration on the blue side was
achieved with a HeHgCd lamp; the red side was calibrated with a NeAr lamp. The
extraction aperture was 4\arcsec \ along the cross-dispersive axis. Spectra
were then flux calibrated and extinction corrected with a standard star closely
matching in airmass and time.

On April 11 there was an electronic irregularity with the red-side CCD. Several
exposures were highly contaminated with a banding structure (i.e., entire pixel
rows were forced to the same count level). This was eventually attributed to a loose
connection inside the detector resulting in variable voltage. Consequently,
several red-side exposures on this night had to be discarded. Unfortunately, it
is likely that this irregularity was present in most exposures from both
nights, albeit not at a level severe enough to notice readily. We touch upon this issue
again in Section~\ref{sec:disc:contvar}.

We also obtained simultaneous $B$-band observations with the 1-m Nickel
telescope and $V$-band observations with the 0.76-m Katzman Automatic
Imaging Telescope (KAIT; \citealt{Filippenko01}) on both nights. The Nickel CCD
was used in 2$\times$2 binned mode, resulting in a $1024\times 1024$ pixel
detector with a pixel scale of $0\farcs 37$ pixel$^{-1}$, along with 5 min
exposures. The Nickel data were bias subtracted automatically and flat fielded
using standard IRAF routines. KAIT uses a back-illuminated Apogee $500 \times
500$ pixel CCD with a $0\farcs 8$ pixel$^{-1}$ scale. We used 4 min
exposures. The KAIT data were reduced automatically as described by
\citet{Li01}.

\subsection{Wise Observatory}

$V$-band observations were obtained on 2004 April 9--10 and 10--11, and on 2004
July 3, with the Wise Observatory 1-m telescope in Israel, using a
back-illuminated Tektronix $1024\times 1024$ pixel CCD. The image scale was
$0\farcs 7$ pixel$^{-1}$.  The nucleus was positioned on the northwest corner
of the chip to maximize the number of relatively bright reference stars
included in the field. Exposures of 262 s each were obtained about once per 5
min. One exposure in the $B$ band was also obtained for the purpose of
measuring the \bv colors of the stars in the field.  Observations were carried
out each night from the start of the night until the airmass of NGC\,4395
exceeded $\sim 2$. Typical image quality was 2--3$\arcsec$ FWHM, deteriorating
toward the end of each night at high airmass. Standard debiasing and flat
fielding were performed.

\subsection{Kitt Peak National Observatory}

Spectroscopic observations at the KPNO Mayall 4-m telescope were obtained on
2004 April 10 and 11. We used the Ritchey-Chr\'{e}tien spectrograph with the
KPC-17B grating (527 l mm$^{-1}$) and the TEK2KB detector. Observations with
this setup covered the range of about 4100--7300 \AA\ with a resolution of
4.2~\AA .  We obtained only 6 and 16 exposures of 5 min duration on April 10
and 11, respectively, a result of rain. We elect to
present only the April 11 data, which consist of a longer, nearly uninterrupted
span of observations. The weather for the April 11 data was clear (nearly photometric) until dawn.

The observation and calibration method we used is the same as described by
\citet{Maoz94} and \citet{Kaspi00}. This method achieves spectrophotometric
calibration by rotating the long spectrograph slit to the appropriate position
angle (PA) so that a nearby comparison star is observed simultaneously with the
object. The comparison star we used is located at a PA of 51\deg \ relative to
the nucleus of NGC\,4395 and at a distance of 33$\arcsec$ (star 1 in
Figure~\ref{fig:field}). We used a slit width of 2$\arcsec$. This PA is only 10\deg \ off the
parallactic angle for this object at high airmass ($\sim 2$), and since the
meridian transit occurs near the zenith, atmospheric dispersion losses for most
of the night are small. We also used the atmospheric dispersion corrector 
(i.e., Risley prisms) to minimize the dispersion losses at this nonparallactic position angle. We expand upon this issue in Section~\ref{sec:spec}.

The spectroscopic data were reduced using standard IRAF routines, and
for each exposure we obtained an AGN/star flux-ratio spectrum. These
spectra were compared to each other to enable cleaning of cosmic-ray 
events. The spectra were calibrated to an absolute flux level using
observations of the spectrophotometric standard star HZ44 taken
adjacent to the observations of NGC\,4395. Although the absolute flux
calibration has an uncertainty of $\sim$10\% , the relative calibration
accuracy between the observations is about 1\% (confirmed
using the \oiii \ line fluxes).

\section{DATA ANALYSIS}
\label{sec:analysis}

\subsection{Photometry}
\label{sec:phot}

\begin{figure}[p]
\plotone{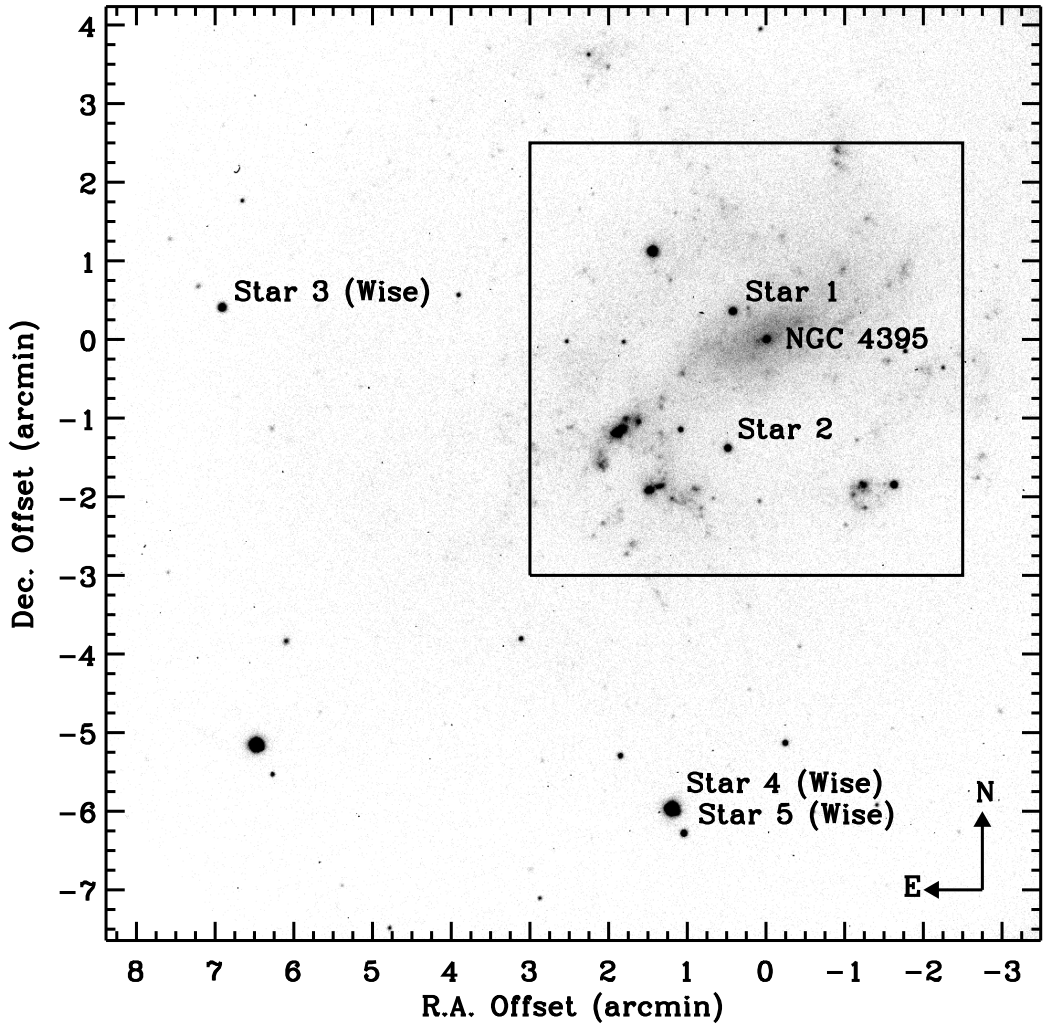}
\caption{The NGC\,4395 field (from the Wise Observatory). The KAIT and Nickel
field is shown as the inner box. The reference stars are labeled. Stars 1 and 2
are common to all three data sets. }
\label{fig:field}
\epsscale{1.0}
\end{figure}

For the KAIT and Nickel data, we examined the images to determine a FWHM for
each night of observations, and found it to be roughly constant at $\sim$4
pixels ($\sim$3\arcsec) for KAIT and $\sim$8 pixels ($\sim$3\arcsec) for Nickel. We chose an aperture with a radius set to this size and performed
aperture photometry on the nucleus and nearby reference (comparison) stars. The sky was subtracted using a 5-pixel wide annulus with an inner radius 20 pixels
from the center. We also attempted point-spread function (PSF) photometry, but
found the results very similar with a slight increase in noise, possibly due to
the relatively faint reference stars, and decided to proceed with aperture
photometry instead. To correct for extinction, non-photometric conditions, and
variable seeing, we normalized the NGC\,4395 flux to those of stars 1 and 2
(see Figure~\ref{fig:field}). These stars were chosen for their proximity to
the nucleus, their similar \bv \ colors, and overlap with the reference stars used in
the analysis of the Wise data. Because of the relative photometry, the
uncertainties in the photometry were estimated using the observed scatter of
the reference stars.

For the Wise data, we chose five stars, all brighter than the nucleus (by 0.4
to 3.6 mag in $V$), with \bv \ colors similar to that of the nucleus (from 0.1
mag bluer to 0.3 mag redder), to serve as photometric reference stars (see
Figure~\ref{fig:field}). Aperture photometry within a fixed, 5-pixel (3\farcs5) radius
aperture was performed on the nucleus and on the stars, and the background
level was determined and subtracted using a 5-pixel wide annulus separated by a
4-pixel interval from the central aperture. This separation is sufficient to exclude the wings of the PSF in the sky annulus. 

The photometric zero point of every exposure was found by taking, for each
star, the instrumental magnitude difference between the exposure and the first
epoch of the night, and calculating the mean difference for all five stars. The
light curve of each star, relative to the zero point determined by the
remaining four stars, was examined and verified to be constant, and its
root-mean-square (rms) fluctuations were noted; these provide an empirical estimate
of the accuracy of the photometry. Outlier points in each star's light curve
were then excluded. The mean of each star's magnitude was subtracted from its
light curve, to avoid any dependence on the zero point of the flux at the first
epoch. Finally, the photometric zero point based on the five stars was
recalculated, and used to calibrate the measurements of the nucleus.

We used the sum of the fluxes from the common reference stars 1 and 2 (Figure~\ref{fig:field}) to
normalize the light curves from all three telescopes. Photometric accuracy,
based on the reference stars, is generally better than 1--2\%, except toward the
end of each night, when the high airmass starts to have an effect. To obtain an
absolute flux scale, we used the Sloan Digital Sky Survey
SkyServer\footnote{http://cas.sdss.org/astro/en/} to retrieve $g$ and $r$
magnitudes for stars 1 and 2. The corresponding calibrated $B$ and $V$ magnitudes
were then found using the transformations of \citet{Smith02}.

\begin{figure}[p]
\plotone{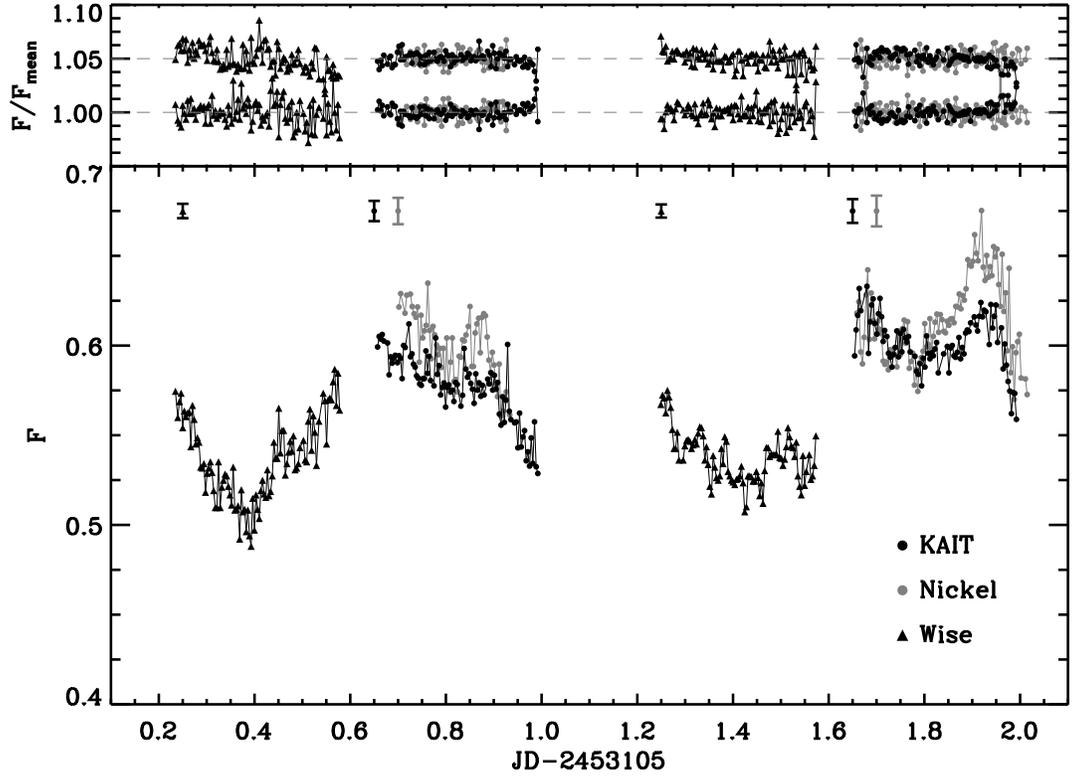}
\caption{Light curves from Wise ($V$ band), KAIT ($V$ band), and Nickel ($B$
band) photometry, 2004 April 10 and 11. The top panel shows, from top to
bottom, stars 1 and 2 (from Figure~\ref{fig:field}), divided by their mean for
each night (plus an offset in the case of star 1, for clarity). The bottom
panel shows NGC\,4395, normalized to the mean flux of stars 1 and
2. Flux is normalized such that $F=1$ corresponds to $V=15.91$ mag and $B=16.69$ mag. Typical
uncertainties are shown for reference above the light curves.}
\label{fig:phot}
\end{figure}

\begin{figure}[p]
\plotone{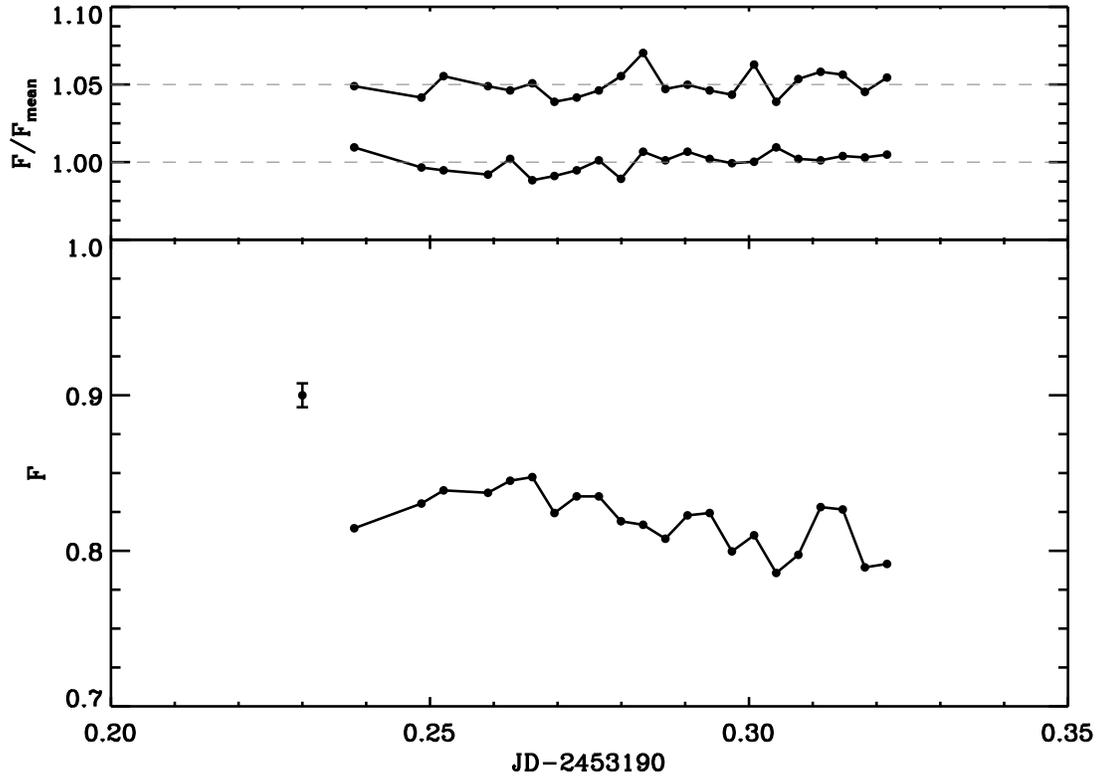}
\caption{Light curve from Wise ($V$ band) photometry, 2004 July 3. The top
panel is as described in Figure~\ref{fig:phot}. The bottom panel shows
NGC\,4395, normalized in the same way as in Figure~\ref{fig:phot}, with
$F$ in the same units. The typical uncertainty is shown for 
reference above the light curve.}
\label{fig:wise3}
\end{figure}

Figures~\ref{fig:phot} and~\ref{fig:wise3} show the various light curves of the
NGC\,4395 nucleus (reproduced in Table~\ref{tab:photflux}) and the reference
stars. Intra-night variability was greatest during the first night of Wise
observations. Table~\ref{tab:photvar} lists variability parameters for these
light curves. The excess variance $F_{\mathrm{var}}$ is as described by
\citet{Rodriguez97} and represents the measured variance above the noise, $R_{\mathrm{max}}$ is the maximum-to-minimum flux ratio,
and $A_{\mathrm{peak}}$ is simply the peak-to-valley amplitude divided by the
unweighted mean of the night. On April 10, NGC\,4395 was observed for 1.5 hours
longer from KAIT than from the Nickel telescope, when there was a sharp decline
in brightness. It is for this reason that KAIT exhibits a greater
$F_{\mathrm{var}}$ on April 10 than Nickel, even though variability is
generally more pronounced toward bluer colors for typical Seyferts. April 11,
with equal KAIT/Nickel coverage, is a better indicator of relative $V$-band and
$B$-band variability.
 
Figure~\ref{fig:wise3} shows the 2004 July 3 observations (corresponding to
{\it HST} Visit 3 from Paper I) from Wise Observatory. NGC\,4395 was only
observed for 2 hours, before it was too low in the sky. Consequently, it is
unclear whether the low variation amplitude is intrinsic or the result of the
short observing window. The nucleus was a factor of about 1.5 brighter in the
$V$ band in July, to be compared to a factor of 3.4 in the UV (Paper I). In terms of
the higher variation amplitude at shorter wavelengths during this time of
increased activity, NGC\,4395 is once again behaving like a typical AGN (see
Section~\ref{sec:disc:contvar}).

\subsection{Spectroscopy}
\label{sec:spec}

The Lick spectra were grouped according to blue/red side and night. Within each
group, the spectra were cross-correlated against each other and rebinned to
correct for slight wavelength solution discrepancies, to an accuracy of 0.1
\AA.  The spectra were then flux normalized according to strong narrow emission
lines: \oiiiw \ on the blue side, the sum of \siiw \ and \oiaw \ (equally weighted) on the
red side.  Since the red and blue sides have a very small overlap region, which
is compromised by dichroic light loss, it was not possible to calibrate the red
side to the blue side. To remedy this, we used complete KPNO spectra to
determine a mean \oiii /(\sii + \oi ) ratio, and applied this to the Lick
spectra to calibrate the red side to the blue side. The narrow-line flux
normalization provided good relative calibration, but the absolute flux
calibration of KPNO data was superior. We therefore used the \oiii \ flux to
adjust the Lick data to the absolute scale of the KPNO data.

We then obtained flux measurements of \ha , \hb , \hg , \heiiw , and the individual \oiii \ lines as a means to examine the behavior of
the spectra and the scatter of the narrow lines. The resolution on the blue
side is sufficient to completely separate both \oiii \ lines. The continuum for each line was determined by averaging the continuum on either side of the line. Each line is well separated and the continuum is very flat over such a short range in wavelength. All flux measurements,
including those of narrow lines, were simple integrations from
continuum-subtracted spectra. Note that \ha \ also includes \nii , a narrow and
non-varying line. Continuum measurements were obtained at 3650, 4200, 4550,
5100, 5625, 6200, and 6410 \AA . They are median values over a small window
typically 100 \AA \ wide, except for 5100 \AA \ which is 80 \AA \ wide, 5625
\AA \ which is 50 \AA \ wide, and 6410 \AA \ which is 40 \AA \ wide. These
windows have been verified to be featureless in a high S/N coadded spectrum, and are as large as possible to reduce the noise. Despite this, the continuum light curves remain very noisy, especially towards shorter wavelengths, due to the low overall flux level.

Standard $1\sigma$ errors are calculated for each flux measurement (which include Poisson, readout noise, sky, flat-fielding, and calibration errors). Since all
spectra are normalized via narrow lines, which are assumed to be constant,
there is potential for extra systematic errors. Indeed, we find the rms scatter of \oiii
, \oii , \oi , and \sii \ to be about 3 times greater than the formal $1\sigma$
errors, and so we adjust our measured errors by this factor to represent the true $1\sigma$ uncertainty in the data.

For consistency, we repeated the above line and continuum flux measurements
with the KPNO data (except \oii \ and 3650 \AA, which are outside the wavelength coverage). Since those
spectra were observed and reduced with a different strategy, this provided a
good check on all our measurements. Uncertainties were again estimated using
the scatter of narrow lines, especially \oiii .

\begin{figure}[p]
\plotone{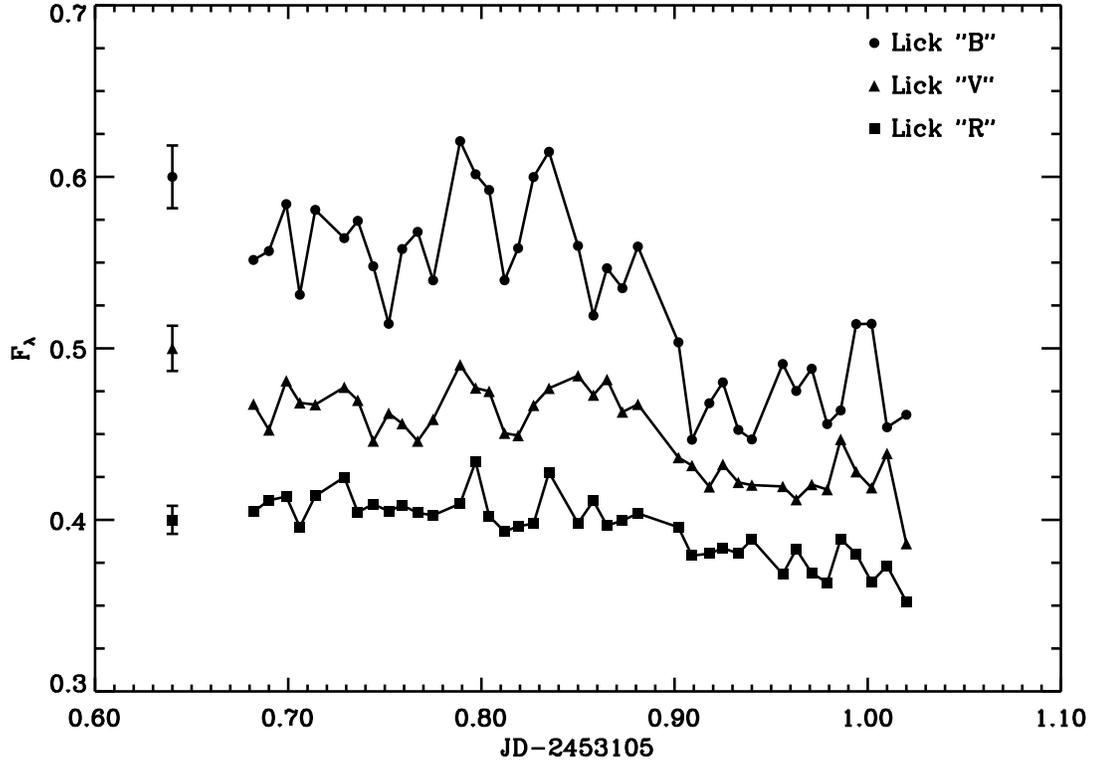}
\caption{Continuum light curves from the Lick data, April 10. Typical
uncertainties are shown at left. The 3650 \AA \ continuum measurement is
extremely noisy and omitted for clarity. The remaining measurements have been paired and averaged to reduce the noise. The Lick ``$B$'' band (circles) is an
average of the 4200 \AA \ and 4550 \AA \ continuum measurements; the Lick
``$V$'' band (triangles) is an average of the 5100 \AA \ and 5625 \AA \
continuum measurements; the Lick ``$R$'' band (squares) is an average of the
6200 \AA \ and 6410 \AA \ continuum measurements. $F_\lambda$ is in units of
10$^{-15}$ \fluxunits.}
\label{fig:lickcont1}
\end{figure}

\begin{figure}[p]
\plotone{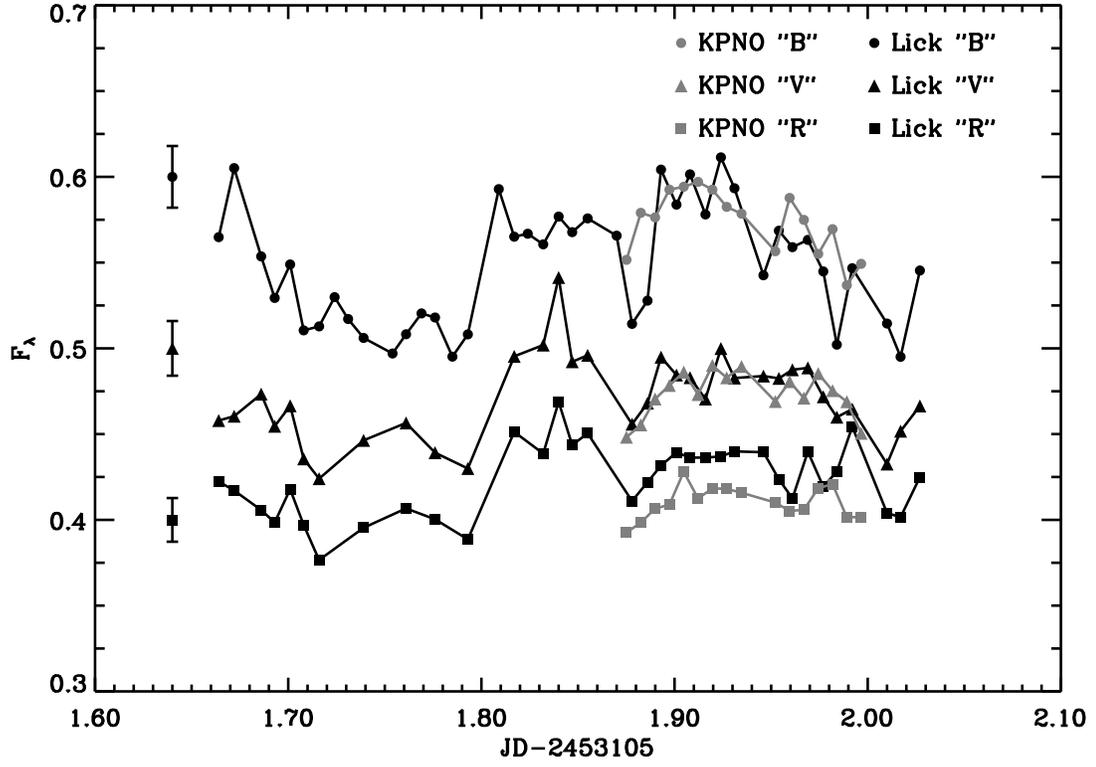}
\caption{Continuum light curves from the Lick data, April 11. Typical
uncertainties are shown at left. The Lick ``$B$,'' ``$V$,'' and ``$R$'' bands
are as described in Fig.~\ref{fig:lickcont1}. KPNO data are also included as
gray symbols, with the ``$B$,'' ``$V$,'' and ``$R$'' bands having the same
definitions as the Lick data. Uncertainties are typically a factor of $\sim$3
smaller than in the Lick data, and are not shown for clarity. $F_\lambda$ is in
units of 10$^{-15}$ \fluxunits.}
\label{fig:lickcont2}
\end{figure}

\begin{figure}[p]
\plotone{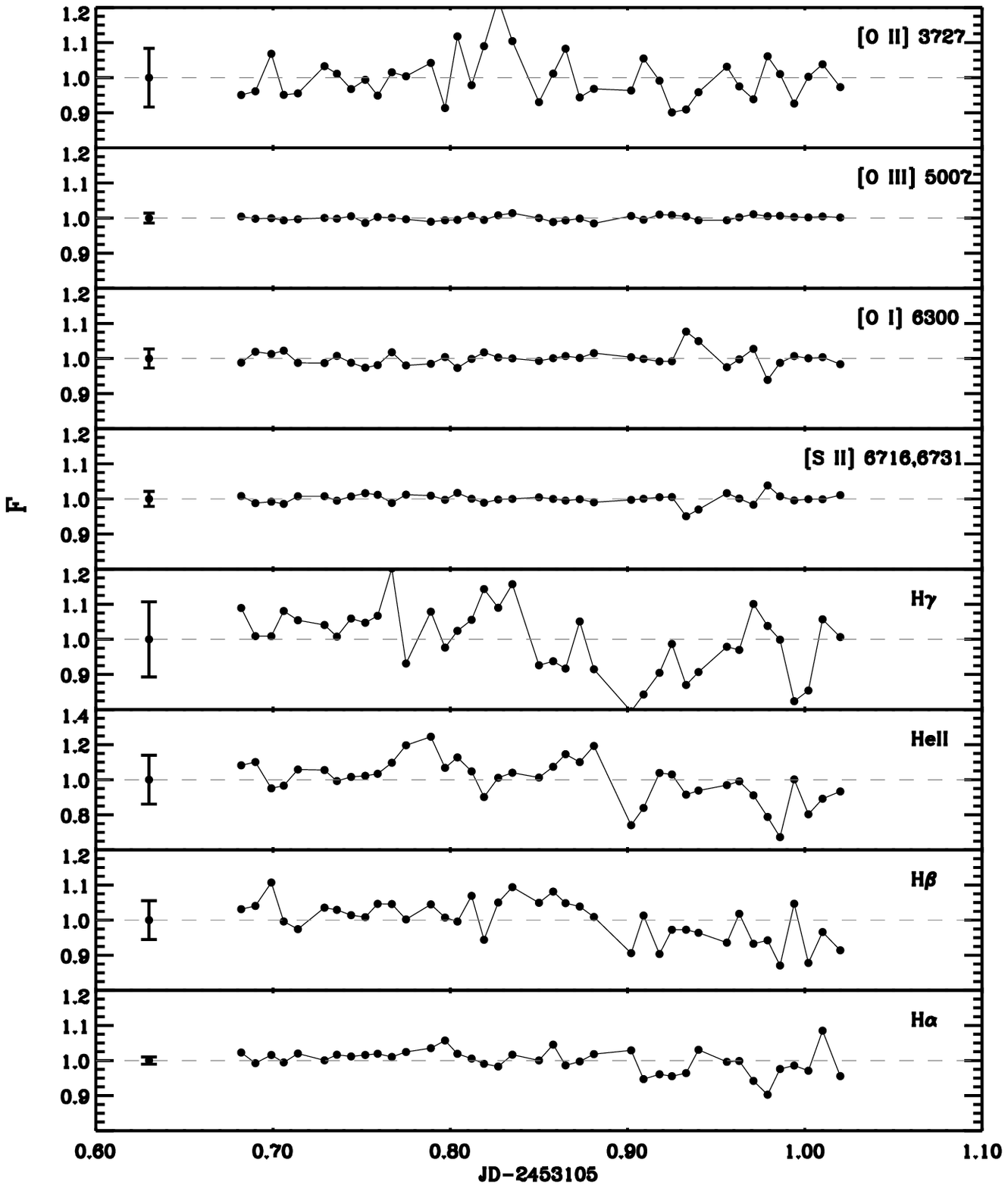}
\caption{Emission-line light curves from the Lick data, April 10. Typical
uncertainties are shown on the left of each panel. Note that \oi \ and \sii \ are mirrored light curves, as their total flux is assumed to be constant. All light curves are normalized
relative to their means; $F$ is therefore in arbitrary units.}
\label{fig:lines1}
\end{figure}

\begin{figure}[p]
\plotone{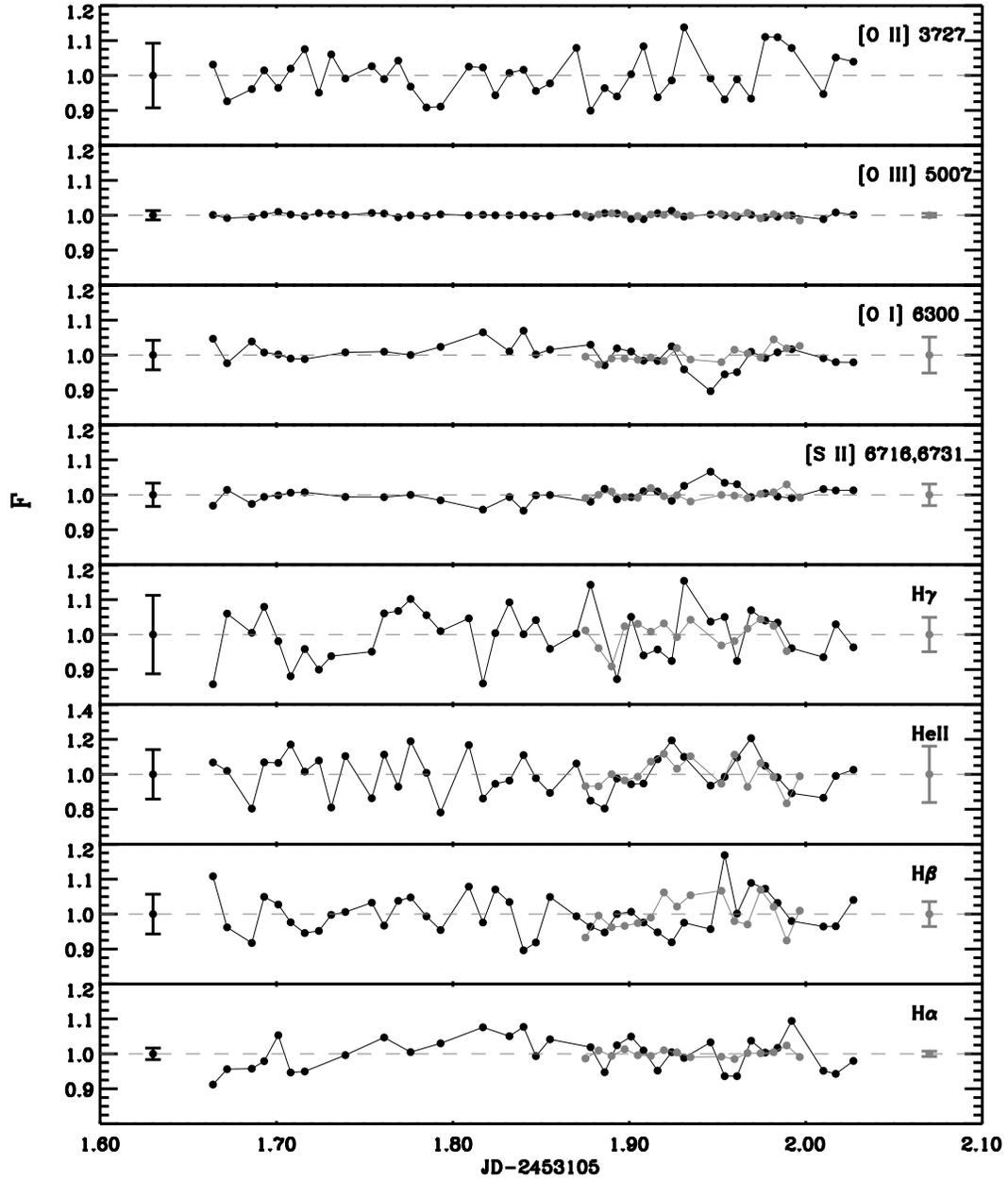}
\caption{Emission-line light curves from the Lick data (KPNO data in gray),
April 11. Typical uncertainties are shown on the left (right) of each panel for
Lick (KPNO) data. Note that \oi \ and \sii \  in the Lick data are mirrored light curves, as their total flux is assumed to be constant. All light curves are normalized relative to their means;
$F$ is therefore in arbitrary units.}
\label{fig:lines2}
\end{figure}

Figures~\ref{fig:lickcont1} and~\ref{fig:lickcont2} show continuum light curves
from the Lick and KPNO data, while Figures~\ref{fig:lines1}
and~\ref{fig:lines2} show the measured light curves of emission lines. These
are also reproduced in Tables~\ref{tab:specflux} and~\ref{tab:lineflux}. In order to decrease the noise, continuum measurements have been paired and averaged, with 3650 \AA \
omitted as it is very noisy. Overall, continuum variability on April 10 is
greater at bluer wavelengths, as expected, while April 11 exhibits optical
variability that is less strongly dependent on wavelength. The agreement
between the Lick and KPNO continuum light curves, despite the different
observing strategies, is reassuring.

The general consistency and stability of the narrow lines on April 10 indicates
no obvious systematic effects, other than increased noise at higher airmass
toward the end of the night. We note that \oi \ and \sii \ are mirror images of
each other, since their sum was used to normalize the spectra. Similarly,
\oiiiaw \ (not shown) is a mirror image of \oiiibw. On April 11, the greater
fluctuations of the narrow forbidden emission lines, especially on the red side
(\oi \ and \sii), are likely a result of slightly compromised
electronics. Weather was also generally worse than on April 10, resulting in lower
S/N data. KPNO measurements are included in Figure~\ref{fig:lines2} for
comparison, with overall agreement between the two sets of data. The major
exception is the \ha \ light curve, which shows substantially more variability
in the Lick data than in the KPNO data. Again, this is likely partially due to
the electronics, with the Lick errors underestimated as a result.

Table~\ref{tab:specvar} lists the variability parameters of the variable emission lines
from the Lick data; we omit the KPNO data due to its short observing
window. The narrow lines have $F_{\mathrm{var}}=0$ by definition (except for \oii, which equals zero due to the noise), and exhibit rms scatter of $\sim$1\% for \oiii, $\sim$2\% for \oi+\sii, and $\sim$6\% for \oii. For comparison, the variability parameters of 1350 \AA \ (corresponding
to {\it HST} Visit 2 on April 11, Paper I) and of the 0.4--8.0 keV X-rays
(corresponding to {\it Chandra} Visits 1 and 2 on April 10 and 11, Paper II)
are also included. For the X-rays, we use 300 s bins and the full duration of each observation ($\sim$30 ks) in order to have similar time-scale ranges. We detect marginal variability in the lines of
interest with broad components. The April 11 \ha \ variability is likely
overestimated, again a result of the electronic difficulties. We detect
significant continuum variability above the noise, which declines toward the red. This is
consistent with photometric results from April 11 (with equal KAIT/Nickel
coverage). The 3650 \AA \ light curve is highly noisy, and no variability
above the noise is detected on April 11. 

It is possible that the non-parallactic KPNO observations (and subsequent Lick
blue-side/red-side calibration using KPNO data) introduced a
wavelength-dependent error, but we estimate such atmospheric dispersion losses
to be minimal given the slit width and PA used.
The use of Risley prisms also minimizes such dispersion losses. 
Inspection of the \oiii /(\sii
+ \oi ) ratio in the KPNO data revealed only a $\sim$4\% discrepancy near the
end of the observations, at high airmass. We did not use these data points,
however, in obtaining a mean ratio to calibrate the blue and red sides of the
Lick spectra. 
Even if dispersion-related
errors persist, we note that this will not affect our measured $F_{\mathrm{var}}$
values, as both the flux and flux error will be erroneous by the same
factor. Only the relative scaling between the blue-side and red-side optical
continua, seen in Figures~\ref{fig:lickcont1} and~\ref{fig:lickcont2}, could be
affected by such errors.

\begin{figure}[p]
\plotone{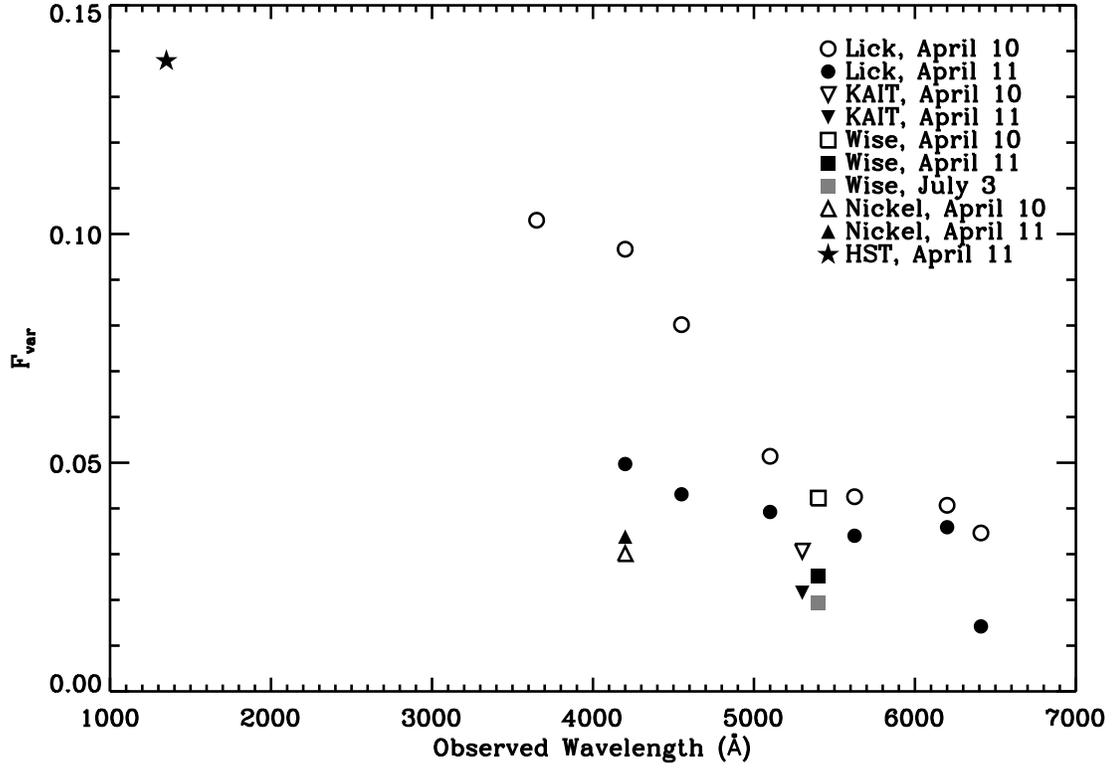}
\caption{Various $F_{\mathrm{var}}$ measurements as a function of wavelength,
from Tables~\ref{tab:photvar} and \ref{tab:specvar}. The ``Lick'' points are
spectroscopic continuum measurements. The {\it HST} point is from Visit 2,
Paper I. The broad-band measurements are placed at the peak of the bandpass,
with KAIT and Wise offset slightly from one another for clarity. Not included
are the values for the 0.4--8.0 keV X-rays from Paper II, which are 0.353 $\pm$
0.020 and 0.362 $\pm$ 0.024 on April 10 and 11, respectively. The important results are (a) the broad trend of increasing variability with decreasing wavelength, (b) the lower overall variability on April 11, and (c) the shallower dependence of variability on wavelength on April 11.}
\label{fig:fvar}
\end{figure}

We plot $F_{\mathrm{var}}$ as a function of wavelength in
Figure~\ref{fig:fvar}. We note that $F_{\mathrm{var}}$ represents measured
variability above the noise. For instance, the Lick
continuum $F_{\mathrm{var}}$ values with smaller windows (5100, 5625, and 6410
\AA ) are somewhat lower than their neighbors as a result of higher
noise. Recall also that the Nickel data on April 10 span a shorter time than
either the KAIT or the Lick 3-m data. What is clear from Figure~\ref{fig:fvar}
is the {\em broad} trend of increased variability with decreasing wavelength, the obviously lower
variability on April 11, and the shallower dependence on wavelength on
April 11.

These $F_{\mathrm{var}}$ measurements are consistent with the general continuum light curve shapes from Figures~\ref{fig:lickcont1} and~\ref{fig:lickcont2} and with the broad trend in spectral slope (i.e., the slope of the continuum). It is quite clear that April 10 exhibits a steepening of the slope with increasing flux (and thus shorter wavelengths have greater variability), whereas on April 11 the slope is nearly constant and a weak function of total flux. This also agrees with the photometric $B-V$ (though on April 10, the Nickel data do not cover the sharp decline in flux at the end of the night, and therefore strong changes in $B-V$ cannot be seen). As the continuum light curves are very noisy, however, especially at short wavelengths, we do not attempt to characterize the spectral slope any further. 

\begin{figure}[p]
\plotone{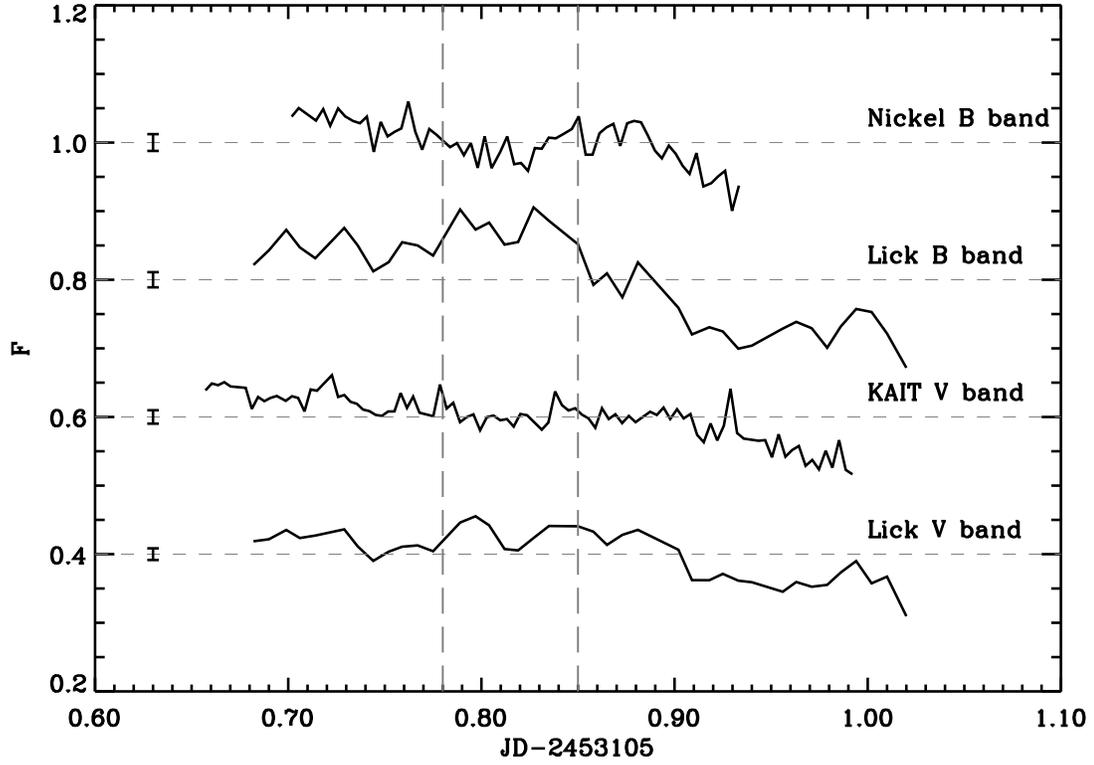}
\caption{Comparison between Lick, KAIT ($V$ band), and Nickel ($B$ band)
data on April 10. KAIT and Nickel data are as described in Fig.~\ref{fig:phot}. Lick
spectra have passed through $B$ and $V$ filter response functions and integrated to create
simulated $B$-band and $V$-band observations. As noted in the text, the Lick
continuum measurements are likely not as reliable as the KAIT and Nickel
measurements, especially during the interval between the vertical dashed
lines. Typical uncertainties are shown at left. All light curves are
normalized relative to their means and shifted vertically for clarity;
$F$ is therefore in arbitrary units.}
\label{fig:lickkaitnick1}
\end{figure}

\begin{figure}[p]
\plotone{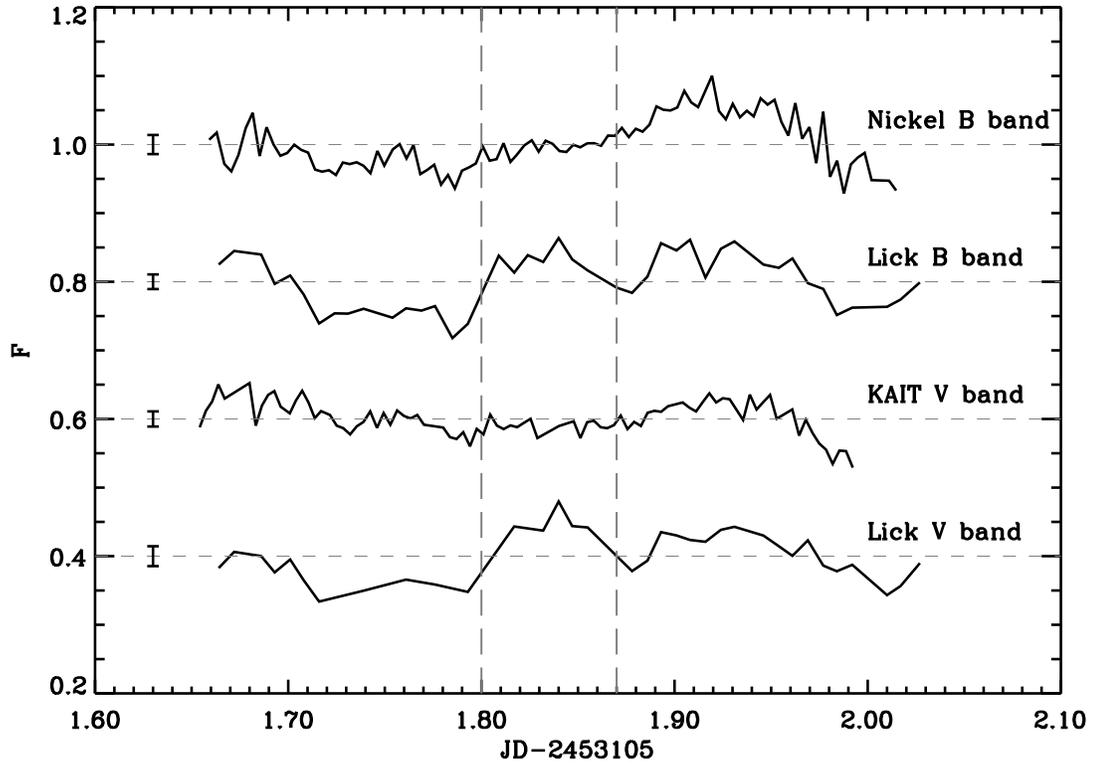}
\caption{Comparison between Lick, KAIT ($V$ band), and Nickel ($B$ band) data on April 11. Light
curves and vertical dashed lines are as described in
Fig.~\ref{fig:lickkaitnick1}. Typical uncertainties are shown at left. All light curves are
normalized relative to their means and shifted vertically for clarity;
$F$ is therefore in arbitrary units.}
\label{fig:lickkaitnick2}
\end{figure}

Figures~\ref{fig:lickkaitnick1} and~\ref{fig:lickkaitnick2} show a comparison
between KAIT, Nickel, and Lick 3-m spectroscopic ``broad-band'' light curves. For this purpose, Lick spectra were passed
through two filter functions  and integrated in order to mimic $V$-band and $B$-band
observations (Table~\ref{tab:specsimflux}). The sharp rise in UV luminosity on April 11 (Figure 1 from Paper I) is far
more subdued in the $B$ and $V$ bands. There is overall agreement between the
two $B$-band and $V$-band data sets to within $\sim$5--10\%. There exists an
odd discrepancy, however, between the Lick spectroscopic ``broad-band'' light
curves and the photometric light curves for a short period on both nights
(JD--2453105, 0.78--0.85 and 1.80--1.87). This discrepancy is mirrored in the
individual spectroscopic continuum light curves from
Figures~\ref{fig:lickcont1} and~\ref{fig:lickcont2}. These occur while
NGC\,4395 crosses the meridian, and correspond to a distinct subgroup of
observations on both nights.

The discrepancy is not due to contamination from an \hii \ region as the slit
rotated over the nucleus, because extra line flux would cause a {\em lower}
continuum with our normalization. Line intensity ratios, such as \oiii/\hb,
also remain constant throughout the night. Galaxy contamination is not the
source either, because (a) the galactic stellar light is very symmetrical about
the nucleus, and (b) galactic stellar light is almost nonexistent in this
bulgeless system \citep{FH03}. We also re-reduced the suspect data with a
different standard star (with similar airmass) and saw little change. Comparing
the calibration coefficients between Lick and KAIT data (which have nearly
identical weather) suggests that the narrow emission-line calibration is indeed
at fault, and not extra continuum light. 

The cause of this calibration anomaly
is unknown. As noted previously, however, an electronic irregularity forced us
to discard several red-side images on April 11. This problem occurred only near
transit, when the position of the telescope was such as to cause the greatest
disruption with the loose cable. Since the suspect data occur near transit, it is
possible that this discrepancy is the result of a subtle electronic glitch
affecting both CCDs, though this is only speculation. Whatever the cause, we
caution the reader about any continuum measurements from the Lick data during these short
periods of time.

\subsection{Cross-Correlations}
\label{sec:xcor}

\begin{figure}[p]
\vspace{2cm}
\plotone{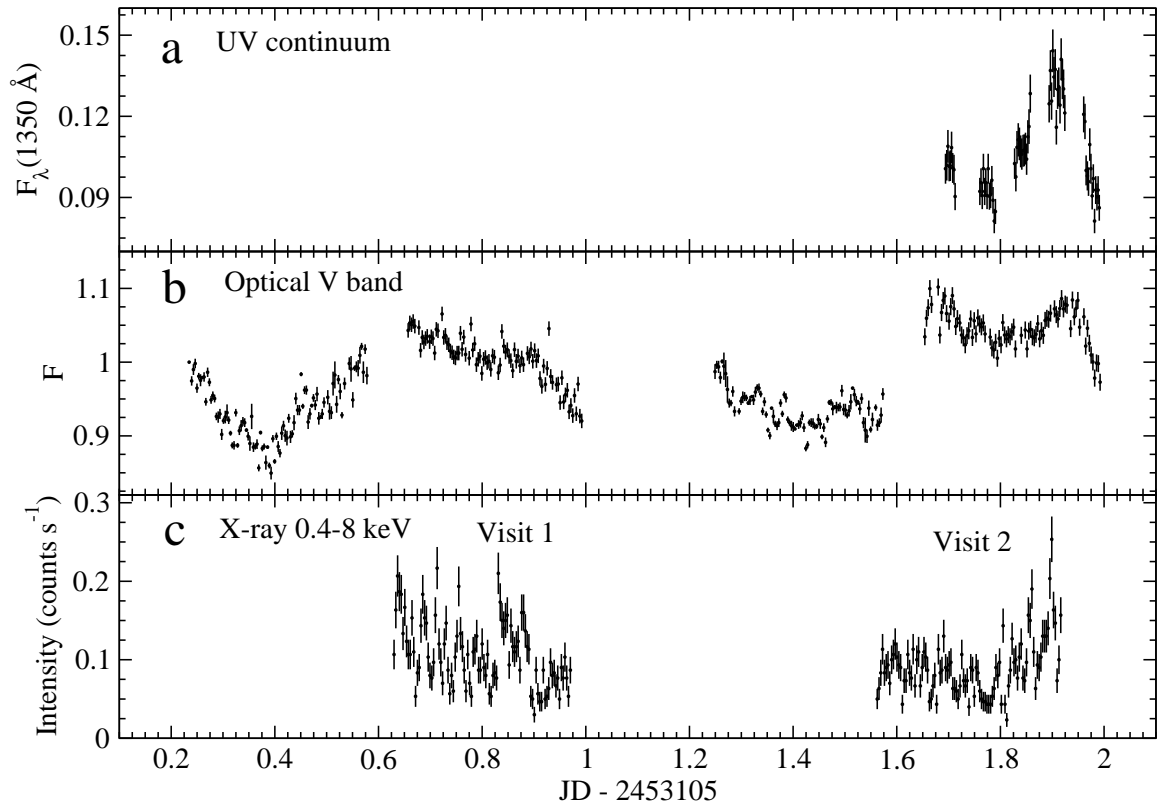}
\caption{Light curves of NGC\,4395 in the three bands used for the
cross-correlation analysis in this paper. (a) UV continuum at 1350 \AA\ binned
to 200 s. Flux in units of $10^{-14}$\,erg\,cm$^{-2}$\,s$^{-1}$\,\AA$^{-1}$
\citep{Peterson05}. (b) $V$-band light curve from Figure~\ref{fig:phot} binned
to $\sim$300 s. Flux in arbitrary units. (c) X-ray light curves presented in the top panels of Figure 1
and 2 of \citet{ONeill06} binned to 300 s.}
\label{fig:multilc}
\end{figure}

Our optical observations were carried out simultaneously with {\it HST} STIS
observations as described in Paper I and with {\it Chandra} ACIS-S observations
which are described in Paper II.  Unfortunately, during the night of April 10,
{\it HST} observations were acquired under gyro control which degraded the
photometric accuracy, rendering these observations useless for our purposes.
We used the April 11 observations along with the two {\it Chandra} observations
for a cross-correlation analysis with the optical observations. The light
curves used in this analysis are presented in Figure~\ref{fig:multilc}. We
refer to the observations of 2004 April 10 and 11 as Visit 1 and Visit 2,
respectively, to be consistent with Papers I and II.

In the following we use two cross-correlation methods: the interpolated
cross-correlation function (ICCF; e.g., \citealt{GS86}; \citealt{GP87}; \citealt{WP94};
\citealt{Peterson04}) and the z-transform discrete correlation function (ZDCF)
of \citet{Alexander97} which is an improvement over the discrete correlation
function (DCF; \citealt{EK88}).\footnote{The ZDCF applies Fisher's $z$
transformation to the correlation coefficients, and uses equal population bins
rather than the equal time bins used in the DCF.}

To calculate the uncertainties in the cross-correlation lag determination we
used the model-independent FR/RSS Monte Carlo method of
\citet{Peterson98,Peterson04,Peterson05}.  In this method, each Monte Carlo
simulation consists of two parts. The first is a ``random subset selection''
(RSS) procedure which consists of randomly resampling a new
light curve of $N$ points from the original light curve of $N$ points, without regard to whether a particular point has been selected previously. Thus
some points are selected multiple times and others are not selected at all. The
uncertainty of a single point that is selected $M$ times is decreased by a
factor of $M^{1/2}$. The second part is ``flux randomization'' (FR) in which
the observed fluxes are altered by random Gaussian deviates scaled to the
uncertainty ascribed to each point. 

The two resampled and altered time series
are then cross-correlated using the ICCF method and the centroid of the CCF is
computed.  We used $\sim$10000 Monte Carlo realizations to build up a
cross-correlation centroid distribution (CCCD; e.g., \citealt{MN89}). The mean
of the distribution is taken to be the time lag and the uncertainty is
determined by the range which contains 68\% of the Monte Carlo realizations in
the CCCD and thus would correspond to 1$\sigma$ uncertainties for a normal
distribution.

\begin{figure}[p]
\vspace{2cm}
\centerline{\includegraphics[scale=0.7]{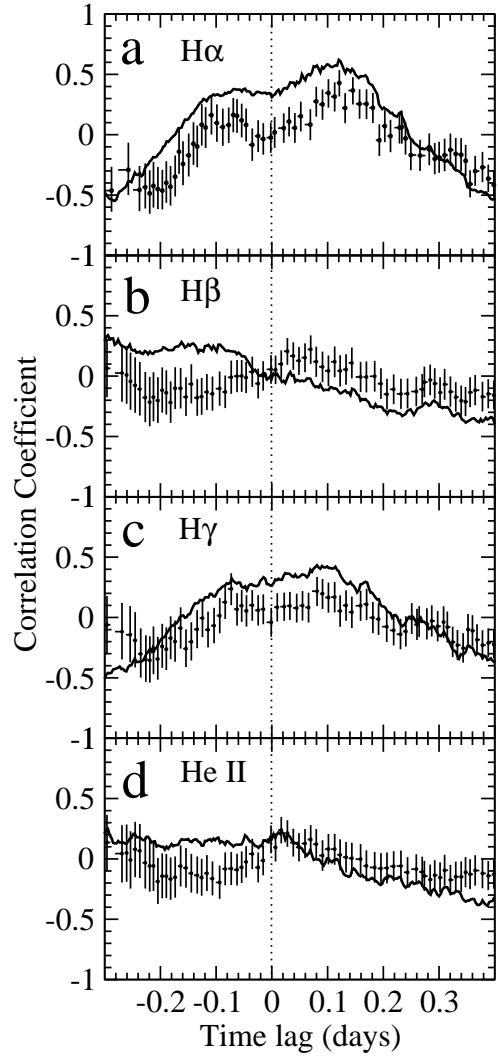}}
\caption{CCFs of the optical-line light curves from Figures~\ref{fig:lines1}
and \ref{fig:lines2} with the optical $V$-band light curve
(Figure~\ref{fig:multilc}b). We present results from the ICCF method (solid
line) and from the ZDCF method (points with error bars). See text for details about comparing these two methods.
(a) Cross-correlation of the \ha \ light curve with the optical light
curve. (b) Cross-correlation of the \hb \ light curve with the optical light
curve. (c) Cross-correlation of the \hg \ light curve with the optical light
curve.  (d) Cross-correlation of the \heiiw \ light curve with the optical
light curve.}
\label{fig:lineccfs}
\end{figure}

Cross-correlation functions of the emission-line light curves from
Figures~\ref{fig:lines1} and~\ref{fig:lines2} with the $V$-band continuum light
curve (Figure~\ref{fig:multilc}b) are presented in Figure~\ref{fig:lineccfs}. In each case, the emission-line and $V$-band light curves from both nights are combined into one.
The cross-correlation of the \ha \ line light curve yields a time lag of the
line behind the continuum of $0.060^{+0.034}_{-0.030}$~d
(Figure~\ref{fig:lineccfs}a), though $\sim$10\% of the FR/RSS Monte Carlo
simulations failed to yield successful centroid calculations as no well-defined
peak was found.  We note that the electronic difficulties on the red side exaggerate the \ha \ variations but leave the overall shape intact. The cross-correlation of the \hb \ line light curve yields a
very shallow peak using the ZDCF method and no peak at all when using the ICCF
method (Figure~\ref{fig:lineccfs}b).  Indeed $\sim$80\% of the FR/RSS Monte
Carlo simulations failed to yield successful centroid calculations. The
$\sim$20\% successful simulations yield a time lag of the line behind the
continuum of $0.047^{+0.050}_{-0.051}$ d.  The cross-correlation the \hg \ line
light curve has a more clearly defined peak than for \hb \ though here also
$\sim$40\% of the FR/RSS Monte Carlo simulations failed to yield successful
centroid calculations. The $\sim$60\% successful simulations yield a time lag
of the line behind the continuum of $0.059^{+0.057}_{-0.060}$ d
(Figure~\ref{fig:lineccfs}c).  The FR/RSS Monte Carlo simulations for the
\heiiw \ line failed in $\sim$70\% cases, again indicating that the CCF does
not have a well-defined peak. The $\sim$30\% successful simulations yield a
time lag of the line behind the continuum of $0.043^{+0.069}_{-0.064}$ d
(Figure~\ref{fig:lineccfs}d). 

We repeated the above cross-correlations with the emission-line light curves
without the problem transit points described in Section~\ref{sec:spec}, to
check for any possible systematic errors. Our results were consistent, with any
variation clearly within the noise of our simulations, and so we proceeded with
our original cross-correlations above. This provides assurance that the problems associated with the anomalous continuum measurements during transit are not propagated to the emission-line light curves.

\begin{figure}[p]
\vspace{2cm}
\centerline{\includegraphics[scale=0.7]{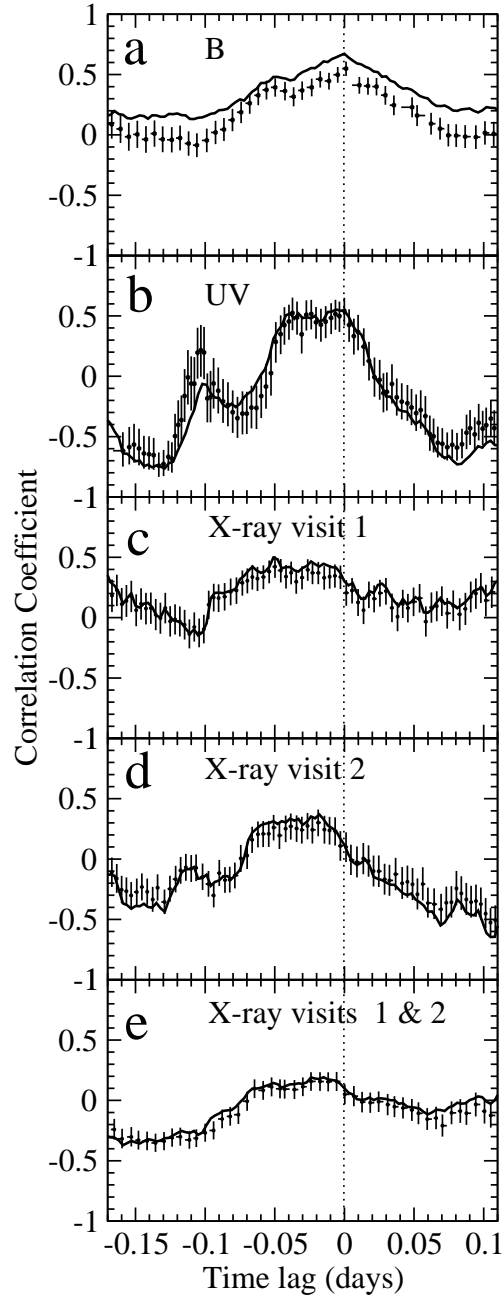}}
\caption{CCFs of the $B$-band, UV, and X-ray continuum light curves with the optical
$V$-band light curve, as presented in Figure~\ref{fig:multilc}. We present
results from the ICCF method (solid line) and from the ZDCF method (points with
error bars).  The results from both methods are consistent. See text for
details.  (a) Cross correlation of the $B$-band
light curve (Figure 2) with the $V$-band light curve (Figure 11b). (b) Cross-correlation of the UV light curve
(Figure~\ref{fig:multilc}a) with the optical light curve
(Figure~\ref{fig:multilc}b). (c) Cross-correlation of the X-ray data from Visit
1 only (Figure~\ref{fig:multilc}c left part) with the simultaneous optical
light curves (Figure~\ref{fig:multilc}b left part). (d) Cross-correlation of
the X-ray data from Visit 2 only (Figure~\ref{fig:multilc}c right part) with
the simultaneous optical light curves (Figure~\ref{fig:multilc}b right part).
(e) Cross-correlation of the X-ray data from Visits 1 \& 2
(Figure~\ref{fig:multilc}c) with the simultaneous optical light curve
(Figure~\ref{fig:multilc}b).}
\label{fig:ccfs}
\end{figure}

We note that the ICCFs and ZDCFs presented in Figure~\ref{fig:lineccfs} are only loosely consistent with each other. In particular, the \hb \ line (Figure~\ref{fig:lineccfs}b) exhibits the greatest difference between the two correlation methods. This arises because of the very small variation amplitude in the underlying light curves and the fact that there is a large gap between the two nights. This gap is interpolated over when using the ICCF method, which produces some extra correlation that distorts the ICCF from the ZDCF. It is reasonable to assume that in this case, the ZDCF is a better representation of the true CCF. Since the correlations and the time lags are of very low significance, however, and are only suggestive, we did not investigate this issue further. The CCFs in Figure~\ref{fig:ccfs} do not show this problem because the underlying light curves have higher variation amplitudes. 

All CCFs presented in Figure~\ref{fig:lineccfs} therefore have some, but
not very significant, indication of a positive time lag behind the optical
continuum.  Although individual results are either consistent with zero or of low significance, the trend as a whole is quite suggestive. In particular, the average lag of the Balmer emission lines is about $0.056\pm0.047$ d, or $80\pm68$ min.  This time lag, though highly
uncertain, is about a factor of 1.5 larger than the time lag found for the \civ
\ line in Paper I. It is also consistent with the $R_{\mathrm{BLR}}$-luminosity relations of \citet{Kaspi00,Kaspi05},
which suggest that the Balmer line-emitting region is of order light hours in
radius in NGC\,4395.

In Figure~\ref{fig:ccfs} we present the cross-correlation functions of the $B$-band, UV,
and X-ray light curves with the optical $V$-band light curve. Note that a negative time lag implies the $V$-band light curve is lagging behind the other reference light curve. No formal time lag is detected between
the $B$-band and $V$-band light curves, with a CCF centroid of $-0.0064^{+0.0074}_{-0.0076}$ d (Figure~\ref{fig:ccfs}a). Our results suggest a 
time lag between the UV and optical light curves of
$-0.017^{+0.006}_{-0.005}$ d (Figure~\ref{fig:ccfs}b).  Within the 1$\sigma$ uncertainty quoted here
this is not consistent with zero time lag and implies the optical emission is
lagging behind the UV emission by $24^{+7}_{-9}$ min. 

For the cross-correlation of the X-ray and optical light curves, we first used
each of the two X-ray visits alone with the simultaneous optical (KAIT)
observations and then cross-correlated both visits as one light curve with the
corresponding optical light curve. These CCFs are presented in
Figures~\ref{fig:ccfs}c--e.  Cross-correlating only the X-ray light curve from
Visit 1 with the optical light curve yields a time lag of
$-0.032^{+0.009}_{-0.009}$ d (Figure~\ref{fig:ccfs}c).  Cross-correlating only
the X-ray light curve from Visit 2 with the optical light curve yields a time
lag of $-0.035^{+0.009}_{-0.007}$ d (Figure~\ref{fig:ccfs}d).
Cross-correlating the X-ray light curve from both visits with the optical light
curve yields a time lag of $-0.026^{+0.012}_{-0.009}$ d
(Figure~\ref{fig:ccfs}e). 

We note that due to the shape of the optical light curve during Visit 1, being
generally monotonically decreasing, the peak seen in the cross-correlation with
the X-ray data (Figure~\ref{fig:ccfs}c) is shallow and broad. In Visit 2 more
structure is seen in the optical light curve, resulting in a somewhat better
defined peak when cross-correlating the light curves (Figure~\ref{fig:ccfs}d).  The strong and fast
variations evident in the X-ray data are not seen in the optical light curve,
which also causes the peak cross-correlation coefficient to be relatively small
(of order 0.3--0.4).  Thus the cross-correlation is more an indication of the
general trend of the optical light curve seen in the X-ray light curve.  We
also note that the average optical flux during Visit 2 was higher than in Visit
1, but in X-rays both visits have about the same flux. This contributes to
the very small peak correlation coefficient seen in Figure~\ref{fig:ccfs}e,
when cross-correlating the data from both visits at once.  Nevertheless, even
with the small correlation coefficients, by simply averaging the above three
results we can determine that the optical continuum is lagging the X-ray
continuum by about $0.031\pm0.009$ d, or $44\pm13$ min.

Cross-correlation of the X-ray light curve with the UV light curve is presented
in Paper II, where the time lag is found to be $-0.003\pm0.016$ d, consistent
with zero lag, with a 95\% upper limit of roughly $\pm0.04$ d, or $\pm1$ hour. We note that the reason for not being able to
firmly determine a time lag between the UV and the X-rays is probably because
these light curves do not overlap at the end of Visit 2. Thus the strong
decrease seen at the end of the UV light curve is not seen in the X-ray light
curve. This prevents us from significantly determining a time lag. The large
gaps in the UV light curve due to the {\it HST} orbit also diminish the
possibility of a firm determination of the lag.

Finally, as a consistency check on our observations, we performed cross-correlations between the Lick continuum ``$B$'' and ``$V$'' light curves (Figures~\ref{fig:lickcont1} and~\ref{fig:lickcont2}) with the simultaneous photometric $B$-band and $V$-band light curves (Figure~\ref{fig:phot}). The bad transit points were removed prior to our calculations. In every case but one, the CCFs were clearly peaked at zero lag (as expected) with correlation coefficients ranging from 0.3 to 0.8. The one exception was the $V$-band correlation on April 10. The CCF was very broad and flat, but this is easily understood given the gradually, monotonically decreasing light curve in the $V$ band. By comparison, the $B$ band exhibits more structure, and so a lag is more easily computed.

We also performed cross-correlations between the spectroscopically derived broad-band light curves from Figures~\ref{fig:lickkaitnick1} and~\ref{fig:lickkaitnick2} with their corresponding photometric light curves. The transit points were again removed. We consider these derived broad-band light curves a much better indicator of the spectroscopic data quality than the individual continuum light curves, because far more flux enters the simulated band pass, increasing the S/N. Once again, three of the four correlations were clearly consistent with zero lag, with correlation coefficients ranging from 0.4 to 0.8. The exception was again the $V$-band light curve on April 10. For the same reasons as above, this CCF was very broad and flat and therefore easily understood.

While the Lick spectroscopic data are very noisy, especially in the continuum, these cross-correlation checks above are reassuring. Combined with the KPNO light curves at the end of April 11, we feel confident about the consistency between the various spectroscopic and photometric data sets, and with all our cross-correlations above.

\section{DISCUSSION}
\label{sec:disc}

\subsection{Continuum Variability}
\label{sec:disc:contvar}

Continuum variations in AGNs are generally more pronounced at shorter
wavelengths than at longer ones. This is detected in the photometry of NGC\,4395, especially
between the KAIT and Nickel data on April 11. We see it in the spectra as well,
with all variability parameters increasing from 6410~\AA \ to 3650~\AA. This
trend continues into the UV at 1350~\AA \ (Paper I) and into the 0.4--8.0 keV
X-rays (Paper II). Finally, the overall continuum level shows a larger rise in
the UV than in the optical over the course of three months, from April 11 to
July 3; the $V$-band brightness was 1.5 times larger while the UV was 3.4 times
larger. 

These variability results are in agreement with the broad picture of AGN
variability \citep{Ulrich97}. Previous studies of NGC\,4395, for instance,
revealed the continuum level varying by a factor of $\sim$2.2 at 3800~\AA\ over
six months, as compared to a factor of $\sim$1.3 at 6800~\AA\
\citep{Lira99}. In NGC\,3783, \citet{Reichert94} report a value of
$R_{\mathrm{max}}$ that nearly doubles from 2700~\AA\ to 1460~\AA. In
NGC\,4151, \citet{Kaspi96} found continuum variations of 17\%--35\%
($R_{\mathrm{max}}-1$), increasing from 6925~\AA\ to 4600~\AA. The trend
continues into the near-UV and far-UV. \citet{Edelson96} also measured
$F_{\mathrm{var}}$ values of 24\%, 9\%, 5\%, 4\%, and 1\% in the X-ray, at 1275
\AA, 1820 \AA, 2688 \AA, and 5125 \AA, respectively. In NGC\,5548, one of the
best-studied AGNs to date, \citet{Korista95} found $F_{\mathrm{var}}$ and
$R_{\mathrm{max}}$ to be greatly increasing from 5100 \AA \ to 1145 \AA \
during a simultaneous observing campaign. \citet{Dietrich98} found
$F_{\mathrm{var}}$ to nearly double from the $I$ band to the $B$ band in
3C\,390.3. \citet{Giveon99} analyzed $B$-band and $R$-band light curves of 42 PG quasars
observed during $\sim$ 30--60 epochs over 7 years. They found a rms
variability amplitude of 14\% in $B$ vs. 12\% in $R$. Furthermore
the majority of the quasars, and particularly those that were the
most variable, were significantly bluer when they were brighter.
For example, in all quasars that varied by $\Delta B> 0.2$~mag,
$\Delta B$ was significantly larger than the corresponding $\Delta R$.

As another comparison, \citet{Skelton05} find significant continuum variability
in NGC\,4395 of $\sim$35\% over the course of 3 nights in 1998. They also find
marginal evidence for greater variability in the blue continuum than in the
red. Our results appear to be consistent, if at a lower variability
amplitude. \citeauthor{Skelton05} also attempted to detect, without success,
variability in broad \hb \ by decomposing the line into narrow and broad
Gaussian components. We initially attempted the same procedure in all of the
lines, in order to isolate the highly variable broad component, but found the
results too unreliable. A Gaussian model with one or two components (or four in
the case of the \ha \ complex) is simply an approximation, especially for the
broad lines, that may not be uniquely fit. We indeed found some slight inverse
correlations between various model components, suggesting an unstable solution
to the line profile as a function of time. For this reason we abandoned this
method and proceeded with straightforward line integration of all the lines,
including single-component narrow lines. This method unfortunately includes
non-varying narrow-line flux in the measurement, diminishing the variability
amplitude, but is far more reliable.

\subsection{X-ray/UV/Optical Correlations}
\label{sec:disc:xcor}

Our optical spectroscopy shows hints of a positive lag of the Balmer emission lines behind the
optical continuum. The lag is consistent with being roughly a factor of 1.5 longer than for the UV lines
and the UV continuum. This is expected, as in most objects studied thus far
the time lag for the UV lines (particularly Ly$\alpha$) was always smaller than
that found for the Balmer lines, by a factor of about 
two (e.g., \citealt{Peterson04}; \citealt{OP02}; \citealt{Peterson91}).

In previous multiwavelength monitoring campaigns of AGNs, the temporal
resolutions were typically insufficient to detect any time lag between X-ray,
UV, and optical continuum variations. Indeed, in NGC\,4151, \citet{Edelson96}
place an upper limit of any X-ray/UV/optical interband delay of 0.15~d, while
\citet{Kaspi96} detect no evidence of a UV-optical continuum
lag. \citet{Shemmer03} find significant evidence for correlated X-ray/optical
emission with zero lag (within 1 d) in NGC\,4051, though their results suggest
a very complex relationship between X-ray and optical emission, one that is not
simple reprocessing. In the same system, \citet{Peterson00} also find evidence
for correlated X-ray/optical emission, but again the time resolution of their
experiment is insufficient to determine a lag of order days or less. With a
resolution of $\sim$1~d, \citet{Peterson91} find no significant lag between the
UV and optical continua in NGC\,5548. A reanalysis of earlier data results in
no UV-optical lag for NGC\,5548, NGC\,3783, and Fairall 9, while in NGC\,7469
there is a UV-optical lag of $\sim$1--2~d \citep{Peterson98, Collier98}.

Some experiments were successful at measuring an interband delay. For instance,
in NGC\,4051, \citet{Mason02} detect the UV emission lagging the X-rays by
0.14~d. In Ark\,564, \citet{Shemmer01} detect the UV continuum lagging the
X-rays by $\sim$0.4~d, the optical lagging the UV by $\sim$1.8~d, and the
optical lagging the X-rays by $\sim$2~d (though in general the optical and
X-ray light curves are not strongly correlated), all consistent with
reprocessing models. By contrast, no correlations at any lag between optical,
UV, and X-ray bands in NGC\,3516 were found, arguing against the
reprocessing model in this case \citep{MMEN02,Edelson00}.

Our results are quite similar, although with large uncertainties, with a 
UV-optical lag of $24^{+7}_{-9}$ min and
a X-ray-optical lag of $44\pm13$ min. A high sampling frequency ($\sim$5 min
observations for optical, 200~s for UV, 300~s bins for X-rays) has allowed us
to make these sub-hour lag measurements. The $B$-band and $V$-band correlation is consistent with zero, though it hints at a possible $V$-band lag behind the $B$-band. Given the proximity of these two bands, a formal lag of zero is not surprising. Although no formal lag was detected
between the UV and X-ray light curves, the fact that the optical continuum lags
the X-ray continuum more than the UV continuum is consistent with the idea of
X-ray photons being reprocessed into UV and then optical photons. The difference between the X-ray-optical lag and the UV-optical lag is also well within the 95\% upper limit of $\pm1$ hour for the X-ray-UV lag established in Paper II.

We can further examine the validity of this model by estimating the energies of
the variable components in the three bands. If the X-rays are indeed the
driving force behind UV and optical ($V$) variability, then we must require
$\Delta L_{\mathrm{X}} \ga \Delta L_{\mathrm{UV}} \ga \Delta L_{\mathrm{opt}}$
for the model to be energetically reasonable. In Paper II we determined the
monochromatic flux variability during Visit 2 was $\sim 2 \times 10^{-13}$
\fluxunitsalt \ at both 1350 \AA \ and 1 keV, using the root of the
unnormalized excess variance (i.e., $F_{\mathrm{var}}$ times the mean). For the
$V$ band, the unnormalized excess variance is roughly $0.023 \times (5 \times 10^{-16})$ \fluxunits . This corresponds to a monochromatic flux variability at 5500 \AA \ of $\sim 6 \times 10^{-14}$ \fluxunitsalt, which satisfies our energy requirement.

It is also interesting to note that this model for continuum variability
extends into the near-infrared $JHK$ bands. \citet{Minezaki06} provide evidence
of near-infrared intraday variability in NGC\,4395 which is correlated with
simultaneous $V$-band observations. Furthermore, the near-infrared light curves
lag behind the $V$-band light curve by $7.2^{+15.8}_{-8.6}$ min, which although
consistent with zero lag, hints at a possible short duration lag, similar to
the $B$ and $V$ bands in this paper.

\subsection{Black Hole Mass Determination}
\label{sec:mbh}

Unfortunately, due to the very low activity level, low variability amplitude,
severe depletion of broad-line flux, and high corresponding noise in the light
curves, we are unable to make a significant reverberation-based mass
measurement of the SMBH in NGC\,4395 to compare with the UV reverberation-based
mass of $M_{\mathrm{BH}} = (3.6 \pm 1.1) \times 10^5 \ \Msun$ from Paper I. The
emission-line light curves for the ``variable'' lines failed to produce
significant time lags behind the continuum light curve, as described in
Section~\ref{sec:xcor}. We are also unable to verify the BLR radius
vs. luminosity relationship determined in Paper I.

\begin{figure}[p]
\plotone{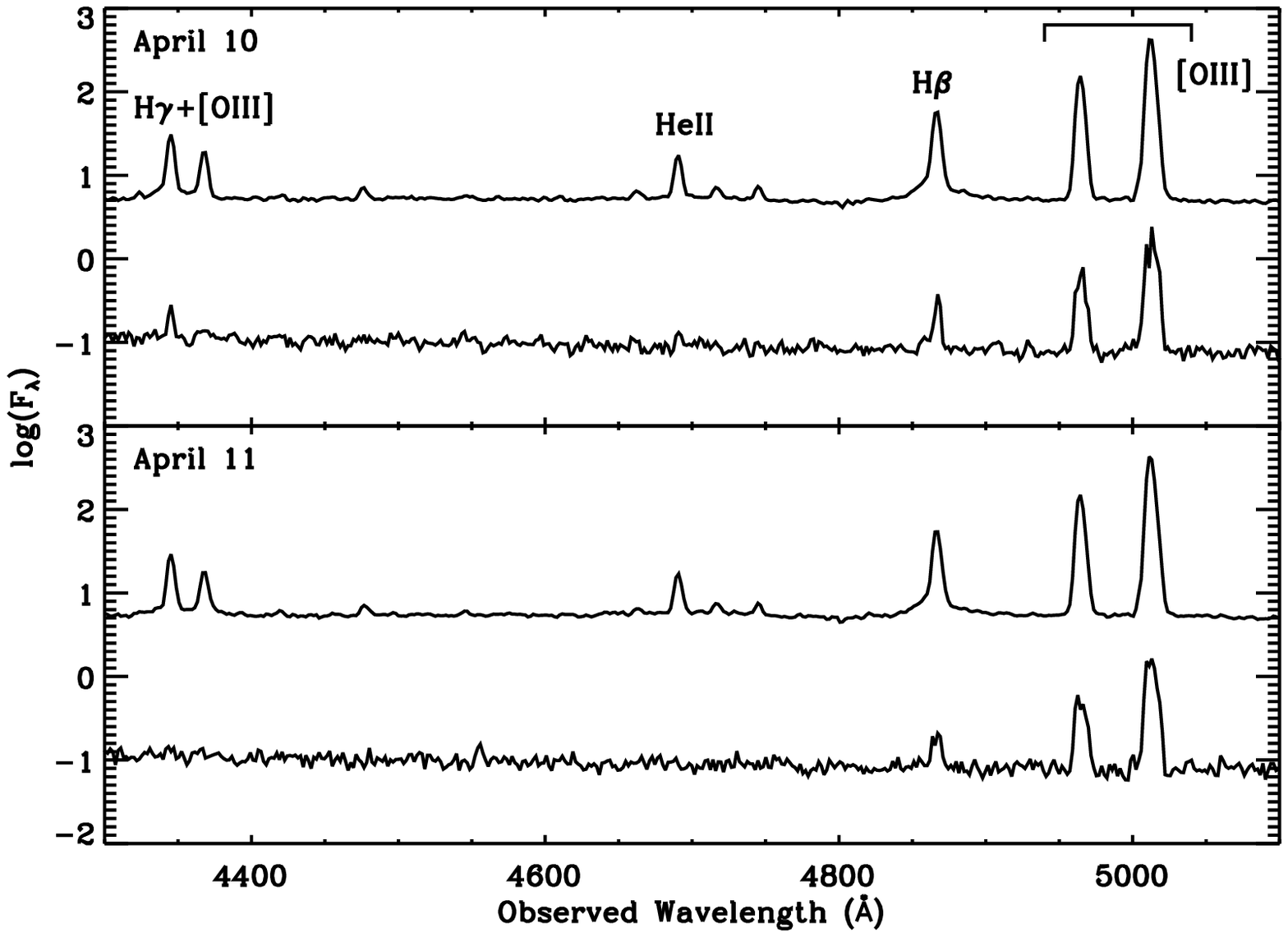}
\caption{Mean and rms blue-side spectra from Lick Observatory. Mean spectra are
plotted above the rms spectra. Mean spectra have been shifted up by 1 dex for
clarity. Overall variability is low, especially in the
broad-line regions. Variable seeing contributes to the narrow lines in the rms spectra. $F_\lambda$ is in units of 10$^{-15}$
\fluxunits.}
\label{fig:rmsb}
\end{figure}

\begin{figure}[p]
\plotone{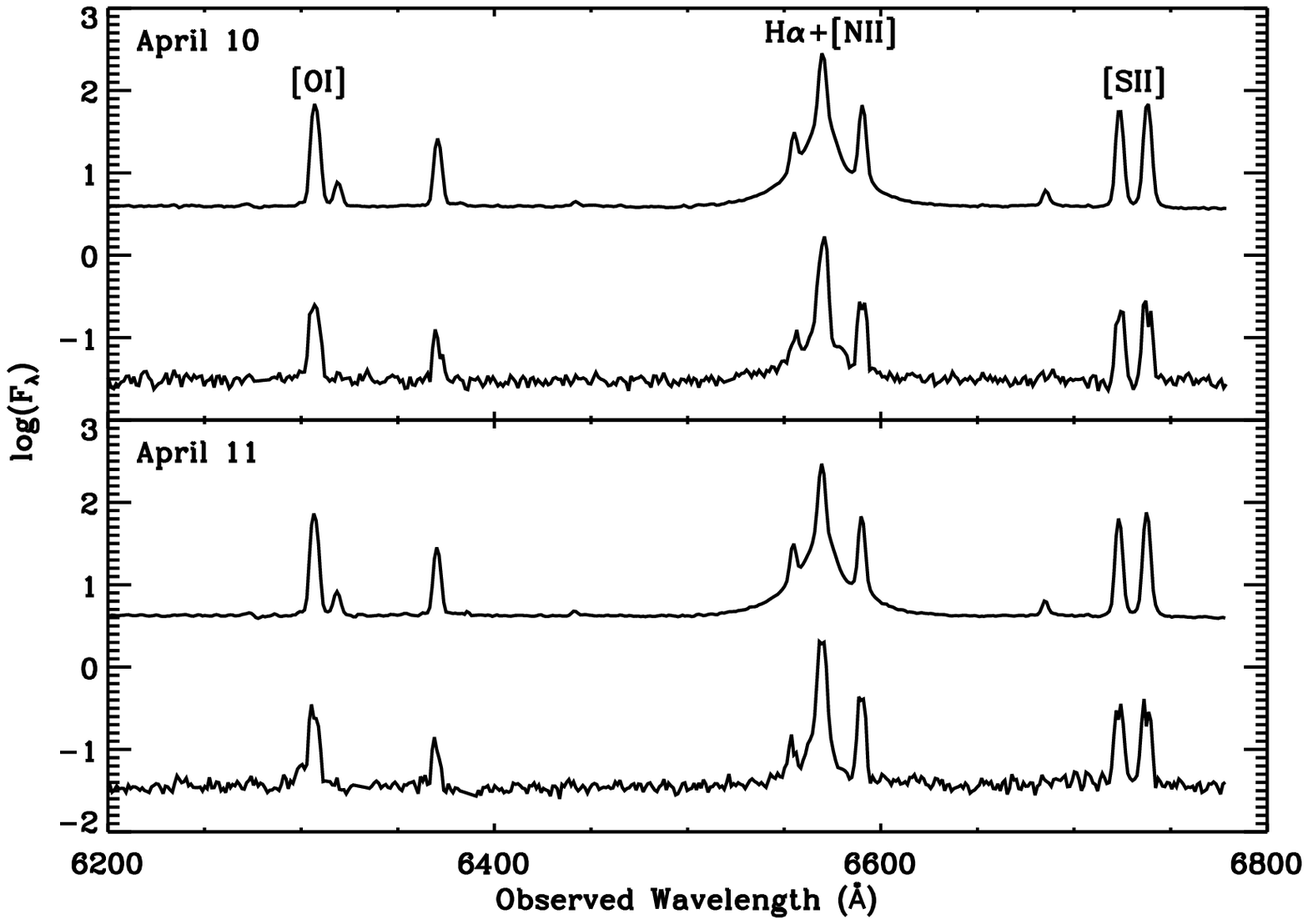}
\caption{Mean and rms red-side spectra from Lick Observatory. Mean spectra are
plotted above the rms spectra. Mean spectra have been shifted up by 1 dex for
clarity. Overall variability is
low. Variable seeing contributes to the narrow lines in the rms spectra. $F_\lambda$ is in units of 10$^{-15}$ \fluxunits.}
\label{fig:rmsr}
\end{figure}

The lack of significant variability can be seen in the rms spectra
(Figures~\ref{fig:rmsb} and~\ref{fig:rmsr}). \heii \ completely  
disappears in
the rms spectra, and broad \hb \ is very weak. The variability in the  
narrow lines is a result of variable seeing. Since the narrow lines  
are not spectrally resolved by the instrument and the seeing disk is smaller  
than the slit width, changes in seeing will alter the line shape as a  
function of time, while keeping the total integrated flux constant.   
Because the resolution of the setup we used with the Lick Kast spectrograph 
is higher on the red side than on the blue side, these line-shape  
variations are more prominent in the red-side rms spectra. 
(On the red side, the noise
induced by the compromised electronics might also be responsible for the
slightly increased rms signal in the narrow lines.)
The broad  
lines are fully resolved, however, so any variability seen in the  
rms spectra is intrinsic. 

\begin{figure}[p]
\plotone{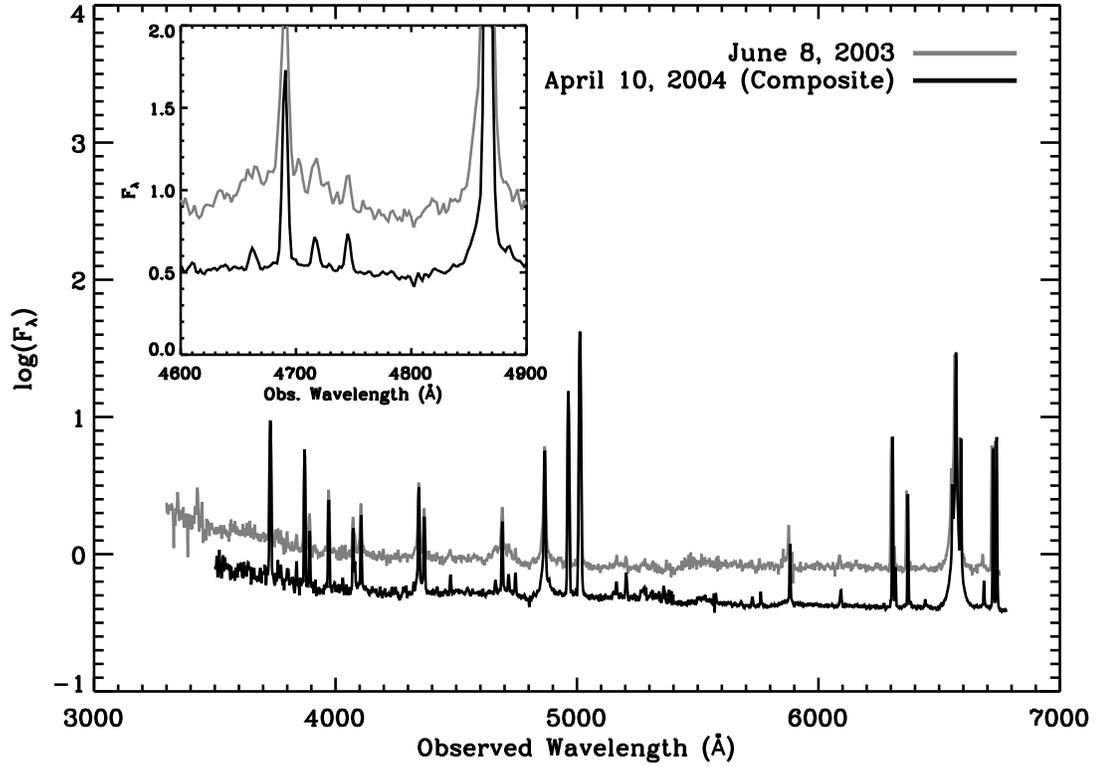}
\caption{Comparison between 2004 April 10 data and 2003 June 8 data, both
obtained at Lick Observatory. Data have been scaled so that \oiiibw \ fluxes
are equal. The inset is a blowup of the \heii \ and \hb \ region. The June 8
spectrum is a single 20-min exposure, while the April 10 composite spectrum has
a total exposure time of 6.58 hr. The latter is clearly at a lower level: the
flux level is lower, the noise (per exposure time) is higher, and the broad-line
components are diminished (especially \heii ). $F_\lambda$ is in units of
10$^{-15}$ \fluxunits.}
\label{fig:epochs}
\end{figure}

The broad component of \ha \ is diminished and the broad \hb \ practically
disappears in the rms spectra. At earlier epochs, the broad \heii \ component
was strong and very broad; it was a prime candidate for reverberation
mapping. As a qualitative comparison, Figure~\ref{fig:epochs} shows a composite
spectrum from 2004 April 10 (total exposure time is 6.58 hours) compared with a
single 20-min exposure taken on 2003 June 8, both from the 3-m Shane telescope
at Lick Observatory. The BLR is noticeably reduced during the low level of
activity in the 2004 data, as is the continuum flux. The two spectra show
similar S/N despite the large difference in exposure times, highlighting the
low flux level in the 2004 data. It should be noted that the 2003 data were
taken with a 4\arcsec \ slit, though as this is an essentially bulgeless
system, there should be insignificant host-galaxy stellar light \citep{FH03}.

Still, our data do show hints of a possible emission-line lag of $80\pm68$ min,
and it is constructive to make a rough estimate of the mass of the SMBH, to
compare with the value derived in Paper I. We estimate the FWHM of the
``variable'' or broad part of \ha, the line exhibiting the greatest
variability, to be $\sim 50$~\AA \ (or $\sim 2300$ km s$^{-1}$). While the
amplitude of the variability may be suspect due to the electronic difficulties,
the lag itself has the greatest significance. Here we assume that the FWHM
$\approx \sigma$ for the broad \ha \ line, as was found for the broad \civ \
line in rms spectra (Table 4 in Paper I). While this is a poor approximation
for a pure Gaussian line, the broad lines are not strictly Gaussian.

The mass of the SMBH is given by
\begin{equation}
M_{\mathrm{BH}} = \frac{f\sigma^2c\tau}{G},
\end{equation}
where $c$ is the speed of light, $G$ the gravitational constant, and $f$ is a
scale factor that depends on the unknown geometry and kinematics of the
broad-line region. To be consistent with Paper I, we use the value found
empirically by \citet{Onken04} of $\langle f \rangle = 5.5$. Using $\tau
\approx 80$ min, we therefore find $M_{\mathrm{BH}} \approx 3 \times 10^5
\Msun$, which is consistent with our results from Paper I. We stress
again that this is simply a qualitative comparison, as our data are of
insufficient quality to provide a significant black hole mass determination.

\section{CONCLUSIONS}
\label{sec:con}

We observed NGC\,4395, the least luminous known Seyfert 1 galaxy, from five different ground-based
optical telescopes over the course of 2 nights in 2004 April, as part of a
coordinated multiwavelength reverberation-mapping campaign designed to measure
the mass of the central black hole. The results of our observations are as
follows.

\begin{enumerate}

\item NGC\,4395 was in a very low state of activity during this period, with optical
continuum flux levels, variability amplitudes, and broad emission-line fluxes
all diminished compared with earlier epochs.

\item Despite this, we detect intrinsic continuum variability, from both
spectra and broad-band photometry, ranging from about 2\% to 10\%
($F_{\mathrm{var}}$). In general, we clearly see evidence for increased
variability toward shorter wavelengths, consistent with other typical
AGNs. This trend extends into the simultaneous UV and X-ray bands. The overall
shape of the optical continuum light curves agrees with the simultaneous UV and
X-ray light curves. Variability was greatest on April 10, while April 11
exhibited a slightly shallower dependence of variability on wavelength.

\item Cross-correlation of the various continuum light curves suggests that
the optical emission lags behind the UV emission by $24^{+7}_{-9}$ min, and the
optical emission lags behind the X-ray emission by $44\pm13$ min. There is no
formal detection of any lag between the UV and X-ray emission, as described in
Paper II, and between the $B$ and $V$ bands. These results are consistent with the reprocessing model, in which
X-ray photons are reprocessed into UV and optical photons. Our high sampling
frequency has allowed us to make these measurements, which have previously been
very difficult to achieve.

\item We only formally detect a lag between the \ha \ light curve and the
$V$-band light curve. Although all other emission-line light curves are
consistent with zero lag, the trend as a whole is quite suggestive.  In
particular, averaging the Balmer-line correlations with the optical continuum
yields an emission-line lag of $80\pm68$ min behind the continuum. While not
very significant, this result is consistent with the relations of
\citet{Kaspi00} for the size of the BLR. The optical emission-line lag is also
about 1.5 times larger than the UV emission-line lag behind the UV
continuum. We have seen this behavior in other AGNs, which indicates NGC\,4395
is behaving typically, albeit faintly, for an AGN.

\item Using the Balmer emission-line lag and the FWHM of the variable
broad-line region, we estimate the mass of the SMBH to be $M_{\mathrm{BH}}
\approx 3\times 10^5 \Msun$. Again, this measurement is not very significant,
and is meant only to compare with the value of $M_{\mathrm{BH}} = (3.6 \pm 1.1)
\times 10^5 \ \Msun$ derived in Paper I. The results are consistent.

\end{enumerate}

\acknowledgements

L.-B.D. would like to thank Ryan Chornock for the many insightful discussions
and suggestions that were of great assistance. We also thank C. M.
Gaskell for helpful, detailed comments. This research is supported by NASA
through {\it HST} grant GO-09818 from the Space Telescope Science Institute,
which is operated by the Association of Universities for Research in Astronomy,
Inc., under NASA Contract NAS5-26555. L.-B.D. is grateful for a
Julie-Payette/Doctoral Fellowship from the Natural Sciences and Engineering
Research Council of Canada and by the Canadian Space Agency. A.V.F.'s group at
UC Berkeley is supported by National Science Foundation (NSF) grant
AST-0307894, and by a gift from the TABASGO Foundation. S.K. is supported in
part at the Technion by a Zeff Fellowship. A.L. acknowledges support from the
Israel Science Foundation (Grant \#1030/04) and from the Norman and Helen Asher
Space Research Institute. M.C.B. is supported by a Graduate Fellowship from the
NSF.  P.M.O. acknowledges financial support from PPARC. KAIT was made possible
by generous donations from Sun Microsystems, Inc., the Hewlett-Packard Company,
AutoScope Corporation, Lick Observatory, the NSF, the University of California,
and the Sylvia \& Jim Katzman Foundation.  We thank the staffs at Lick and Kitt
Peak National Observatories for their help with the observations.


\clearpage

\begin{deluxetable}{ccc}
\tabletypesize{\scriptsize}
\tablewidth{0pt}
\tablecaption{Broad-Band Continuum Light Curves}
\tablehead{\colhead{JD--2453105} & \colhead{$F\tablenotemark{a}$} & \colhead{$\sigma\tablenotemark{a}$}}
\startdata
\multicolumn{3}{c}{Wise ($V$ band)} \\[5pt]
\hline \\[-5pt]
   0.23469 &  0.574 &  0.000 \\ 
   0.23966 &  0.560 &  0.004 \\ 
   0.24307 &  0.569 &  0.004 \\ 
   0.24648 &  0.573 &  0.004 \\ 
   0.24988 &  0.554 &  0.004 \\
   \vspace{-6pt} 
\enddata
\tablenotetext{a}{\ Flux normalized so that
$F=1$ corresponds to $V = 15.91$ mag and $B=16.69$ mag.}
\tablenotetext{b}{\ Data points with relative dates beginning with 85 refer to 2004 July 3 Wise data.}
\tablecomments{Table~\ref{tab:photflux} is published in its entirety in the
electronic edition of the Journal. A portion is shown here for guidance
regarding its form and content.}
\label{tab:photflux}
\end{deluxetable}
\begin{deluxetable}{lccccccccc}
\tabletypesize{\scriptsize}
\tablewidth{0pt}
\tablecaption{Variability Parameters --- Photometry}
\tablehead{ & \colhead{$F_{\mathrm{var}}$} & \colhead{$R_{\mathrm{max}}$} & \colhead{$A_{\mathrm{peak}}$} & \colhead{$F_{\mathrm{var}}$} & \colhead{$R_{\mathrm{max}}$} & \colhead{$A_{\mathrm{peak}}$} & \colhead{$F_{\mathrm{var}}$} & \colhead{$R_{\mathrm{max}}$} & \colhead{$A_{\mathrm{peak}}$} \\ \colhead{Obs.} & \colhead{April 10} & \colhead{April 10} & \colhead{April 10} & \colhead{April 11} & \colhead{April 11} & \colhead{April 11} & \colhead{July 3} & \colhead{July 3} & \colhead{July 3} \\ \colhead{(1)} & \colhead{(2)} & \colhead{(3)} & \colhead{(4)} & \colhead{(5)} & \colhead{(6)} & \colhead{(7)} & \colhead{(8)} & \colhead{(9)} & \colhead{(10)}}
\startdata
Wise ($V$ band) & 0.042 & 1.20 $\pm$ 0.01 & 0.184 $\pm$ 0.011 & 0.025 & 1.13 $\pm$ 0.01 & 0.126 $\pm$ 0.005 & 0.019 & 1.08 $\pm$ 0.01 & 0.075 $\pm$ 0.010 \\
KAIT ($V$ band) & 0.031 & 1.16 $\pm$ 0.02 & 0.144 $\pm$ 0.014 & 0.022 & 1.13 $\pm$ 0.02 & 0.123 $\pm$ 0.016 \\
Nickel ($B$ band) & 0.030 & 1.18 $\pm$ 0.02 & 0.159 $\pm$ 0.017 & 0.034 & 1.18 $\pm$ 0.02 & 0.171 $\pm$ 0.020 \\
\vspace{-6pt}
\enddata
\label{tab:photvar}
\end{deluxetable}
\clearpage

\begin{deluxetable}{ccc}
\tabletypesize{\scriptsize}
\tablewidth{0pt}
\tablecaption{Spectroscopic Continuum Light Curves}
\tablehead{\colhead{JD--2453105} & \colhead{$F_{\lambda}\tablenotemark{a}$} & \colhead{$\sigma\tablenotemark{a}$}}
\startdata
\multicolumn{3}{c}{3650 \AA \ (Lick)} \\[5pt]
\hline \\[-5pt]
 0.682 &   0.87 &   0.07 \\ 
 0.690 &   0.91 &   0.06 \\ 
 0.699 &   0.76 &   0.07 \\ 
 0.706 &   0.71 &   0.07 \\ 
 0.714 &   0.67 &   0.06 \\
 \vspace{-6pt}
\enddata
\tablenotetext{a}{\ Flux density in units of 10$^{-15}$ \fluxunits.}
\tablenotetext{b}{\ Suspect transit data points. See text for details.}
\tablecomments{Table~\ref{tab:specflux} is published in its entirety in the electronic edition of the Journal. A portion is shown here for guidance regarding its form and content.}
\label{tab:specflux}
\end{deluxetable}

\begin{deluxetable}{ccc}
\tabletypesize{\scriptsize}
\tablewidth{0pt}
\tablecaption{Spectroscopic Emission-Line Light Curves}
\tablehead{\colhead{JD--2453105} & \colhead{$F\tablenotemark{a}$} & \colhead{$\sigma\tablenotemark{a}$}}
\startdata
\multicolumn{3}{c}{\oiiw \ (Lick) } \\[5pt]
\hline \\[-5pt]
 0.682 &   0.95 &   0.07 \\ 
 0.690 &   0.96 &   0.08 \\ 
 0.699 &   1.07 &   0.09 \\ 
 0.706 &   0.95 &   0.08 \\ 
 0.714 &   0.96 &   0.07 \\
 \vspace{-6pt}
\enddata
\tablenotetext{a}{\ Flux in arbitrary units, normalized to the mean of each night.}
\tablecomments{Table~\ref{tab:lineflux} is published in its entirety in the electronic edition of the Journal. A portion is shown here for guidance regarding its form and content.}
\label{tab:lineflux}
\end{deluxetable}

\clearpage
\begin{deluxetable}{lcccccc}
\tabletypesize{\scriptsize}
\tablewidth{0pt}
\tablecaption{Variability Parameters --- Spectroscopy (Lick)}
\tablehead{ & \colhead{$F_{\mathrm{var}}$} & \colhead{$R_{\mathrm{max}}$} & \colhead{$A_{\mathrm{peak}}$} & \colhead{$F_{\mathrm{var}}$} & \colhead{$R_{\mathrm{max}}$} & \colhead{$A_{\mathrm{peak}}$} \\ \colhead{Feature} & \colhead{April 10} & \colhead{April 10} & \colhead{April 10} & \colhead{April 11} & \colhead{April 11} & \colhead{April 11} \\\colhead{(1)} & \colhead{(2)} & \colhead{(3)} & \colhead{(4)} & \colhead{(5)} & \colhead{(6)} & \colhead{(7)}}
\startdata
\hg\tablenotemark{a} & 0 & 1.51 $\pm$ 0.23 & 0.41 $\pm$ 0.14 & 0 & 1.34 $\pm$ 0.25 & 0.30 $\pm$ 0.18 \\
\heii & 0 & 1.85 $\pm$ 0.51 & 0.57 $\pm$ 0.22 & 0 & 1.54 $\pm$ 0.29 & 0.43 $\pm$ 0.18 \\
\hb & 0.020 & 1.27 $\pm$ 0.11 & 0.24 $\pm$ 0.08 & 0 & 1.30 $\pm$ 0.10 & 0.27 $\pm$ 0.08 \\
\ha\tablenotemark{b} & 0.033 & 1.20 $\pm$ 0.02 & 0.18 $\pm$ 0.02 & 0.044 & 1.20 $\pm$ 0.03 & 0.18 $\pm$ 0.03 \\
0.4--8.0 keV\tablenotemark{c} & 0.353 $\pm$ 0.020 & & & 0.362 $\pm$ 0.024 & & \\
$F_\lambda$(1350 \AA)\tablenotemark{d} & & & & 0.138 & 1.77 $\pm$ 0.14 & 0.59 $\pm$ 0.08 \\
$F_\lambda$(3650 \AA) & 0.103 & 2.20 $\pm$ 0.51 & 0.71 $\pm$ 0.25 & 0 & 2.04 $\pm$ 0.56 & 0.68 $\pm$ 0.33 \\
$F_\lambda$(4200 \AA) & 0.097 & 1.50 $\pm$ 0.16 & 0.40 $\pm$ 0.09 & 0.050 & 1.35 $\pm$ 0.09 & 0.30 $\pm$ 0.07 \\
$F_\lambda$(4550 \AA) & 0.080 & 1.44 $\pm$ 0.09 & 0.36 $\pm$ 0.06 & 0.043 & 1.29 $\pm$ 0.07 & 0.26 $\pm$ 0.06 \\
$F_\lambda$(5100 \AA) & 0.051 & 1.36 $\pm$ 0.10 & 0.30 $\pm$ 0.07 & 0.039 & 1.28 $\pm$ 0.07 & 0.25 $\pm$ 0.05 \\
$F_\lambda$(5650 \AA) & 0.043 & 1.25 $\pm$ 0.07 & 0.22 $\pm$ 0.05 & 0.034 & 1.33 $\pm$ 0.09 & 0.29 $\pm$ 0.07 \\
$F_\lambda$(6200 \AA) & 0.040 & 1.26 $\pm$ 0.05 & 0.22 $\pm$ 0.03 & 0.036 & 1.25 $\pm$ 0.05 & 0.22 $\pm$ 0.04 \\
$F_\lambda$(6410 \AA) & 0.035 & 1.23 $\pm$ 0.07 & 0.21 $\pm$ 0.05 & 0.014 & 1.24 $\pm$ 0.08 & 0.21 $\pm$ 0.06 \\
\vspace{-6pt}
\enddata
\tablenotetext{a}{\ Includes \oiiicw.}
\tablenotetext{b}{\ Includes \niiw.}
\tablenotetext{c}{\ {\it Chandra} Visits 1 and 2 (Paper II). The parameters are calculated for the full duration of each observation ($\sim$30 ks) in 300 s bins.}
\tablenotetext{d}{\ {\it HST} Visit 2 (Paper I).}
\label{tab:specvar}
\end{deluxetable}

\begin{deluxetable}{ccc}
\tabletypesize{\scriptsize}
\tablewidth{0pt}
\tablecaption{Spectroscopically Derived Broad-Band Light Curves}
\tablehead{\colhead{JD--2453105} & \colhead{$F\tablenotemark{a}$} & \colhead{$\sigma\tablenotemark{a}$}}
\startdata
\multicolumn{3}{c}{Lick $B$ band} \\[5pt]
\hline \\[-5pt]
 0.682 &  1.02 &  0.01 \\ 
 0.690 &  1.04 &  0.01 \\ 
 0.699 &  1.07 &  0.01 \\ 
 0.706 &  1.05 &  0.01 \\ 
 0.714 &  1.03 &  0.01 \\
 \vspace{-6pt}
\enddata
\tablenotetext{a}{\ Flux in arbitrary units, normalized to the mean of each night.}
\tablenotetext{b}{\ Suspect transit data points. See text for details.}
\tablecomments{Table~\ref{tab:specsimflux} is published in its entirety in the electronic edition of the Journal. A portion is shown here for guidance regarding its form and content.}
\label{tab:specsimflux}
\end{deluxetable}

\clearpage

\clearpage

\end{document}